\documentstyle[11pt,epsf,psfig]{article}

\def\Year{\expandafter\eatPrefix\the\year}
\newcount\hours \newcount\minutes
\def\monthname{\ifcase\month\or
January\or February\or March\or April\or May\or June\or July\or
August\or September\or October\or November\or December\fi}
\def\shortmonthname{\ifcase\month\or
Jan\or Feb\or Mar\or Apr\or May\or Jun\or Jul\or
Aug\or Sep\or Oct\or Nov\or Dec\fi}

\def\TimeStamp{\hours\the\time\divide\hours by60%
\minutes -\the\time\divide\minutes by60\multiply\minutes by60%
\advance\minutes by\the\time%
${\rm \shortmonthname}\cdot   \if\day<10{}0\fi\the\day\cdot   \the\year%
\qquad\the\hours:\if\minutes<10{}0\fi\the\minutes$}

\newskip\humongous \humongous=0pt plus 1000pt minus 100pt
\def\caja{\mathsurround=0pt}
\def\eqalign#1{\,\vcenter{\openup1\jot \caja
       \ialign{\strut \hfil$\displaystyle{##}$&$
        \displaystyle{{}##}$\hfil\crcr#1\crcr}}\,}
\newif\ifdtup

\newcounter{eqnumber}[section]
\renewcommand{\theeqnumber}{\thesection.\arabic{eqnumber}}
\def\equn{\refstepcounter{eqnumber}
\eqno({\rm \theeqnumber})
}

\def\tr{\mathop{\rm tr}\nolimits}

\newbox\charbox
\newbox\slabox
\def\s#1{{      
        \setbox\charbox=\hbox{$#1$}
        \setbox\slabox=\hbox{$/$}
        \dimen\charbox=\ht\slabox
        \advance\dimen\charbox by -\dp\slabox
        \advance\dimen\charbox by -\ht\charbox
        \advance\dimen\charbox by \dp\charbox
        \divide\dimen\charbox by 2
        \raise-\dimen\charbox\hbox to \wd\charbox{\hss/\hss}
        \llap{$#1$}
}}

\def\spa#1.#2{\left\langle#1\,#2\right\rangle}
\def\spb#1.#2{\left[#1\,#2\right]}
\def\lor#1.#2{\left(#1\,#2\right)}

\catcode`@=11  

\def\Tr{\, {\rm Tr}}

\def\eps{\epsilon}

\def\pol{\eps}

\def\x#1#2{x_{#1 #2}}

\def\dlips{dLIPS}

\def\lsl{\not{\hbox{\kern-2.3pt $\ell$}}}
\def\ksl{\not{\hbox{\kern-2.3pt $k$}}}
\def\kaps{\left( { \kappa \over 2 } \right)^2 } 
\def\kapss{\left( { \kappa \over 2 } \right)^4 }

\def\Mtree{M^{\rm tree}}

\begin{document}

\begin{titlepage}

\begin{flushright}
hep-th/9911158  \hfill \TimeStamp \\
\hfill  ENS-99/34  \\ 
SWAT-99-241 \\
November, 1999\\
\end{flushright}

\vskip 2.cm

\begin{center}
\begin{Large}
{\bf Counterterms in type I Supergravities}

{\bf }
\end{Large}

\vskip 2.cm

{\large 
David C. Dunbar $^{\sharp,1}$, 
Bernard Julia $^{\star,1,2}$,
Domenico Seminara $^{\star,2}$ 
and Mario Trigiante  $^{\sharp,1}$ 
}

\vskip 0.5cm

$^\sharp${\it Department of Physics, University
    of Wales Swansea, Swansea,  UK }

\vskip .3cm

$^\star${\it Laboratoire de Physique Th\'eorique, 
CNRS-ENS $^{3}$ , Paris, France}
\end{center}

\begin{abstract}

We compute the one-loop divergences of
$D=10$, $N=1$ supergravity
and of its reduction to $D=8$.
We study the tensor structure of the counterterms 
appearing in $D=8$ and
$D=10$ and compare these to expressions 
previously found in the low energy
expansion of string theory.
The infinities have the primitive Yang-Mills tree amplitude 
as a common factor.

\end{abstract}

\vfill
\noindent\hrule width 3.6in\hfil\break
${}^1$Research supported by TMR grant FMRX-CT96-0012 \hfil\break
${}^2$Research supported by TMR grant FMRX CT96-0045 .\hfil\break
${}^3$UMR 8549 du CNRS et de l'Ecole Normale 
Sup\'erieure.\hfil\break
\end{titlepage}

\baselineskip 16pt



\section{Introduction}

Highly extended supergravity theories have long been 
seen as potential theories of quantum gravity
although they were displaced by
superstring theories \cite{GSW} as favored candidates for a ``final theory''.
The (quantum) M theory has 10 dimensional supergravities as  low energy 
limits and might be thought of 
either as  a (non-perturbative) quantum version of eleven dimensional 
supergravity (including its extended solutions) or as a non-perturbative
completion of superstring theories.
The spectacular perturbative ultra-violet finiteness of string theories 
indicates that strings provide a physical regulator of supergravities in 10 
dimensions.  In this
paper we shall explore the ultra-violet behavior of $D=10\, ,\, N=1$
supergravity and its dimensional descendants from a field theory viewpoint.  
In this we are
attempting to ``work-up'' to the quantum theory of gravity rather than
work down from string theory. 

Of the extended supergravities two play a central role. 
Firstly, there is the $D=11$, $N=1$ maximally extended
theory \cite{CJS} which reduces to $N=8$ in $D=4$ \cite{ExtendedSugra}.
In some ways this theory is the ultimate conventional
point-particle field theory.
The one-loop amplitude is potentially 
infinite for $D\geq 8$ although in dimensional regularisation it is
only infinite for $D=8$ (in the dimensional regularisation
prescription one-loop amplitudes in odd dimensions are finite and the
$D=10$ infinity vanishes onshell). We shall work 
entirely within the dimensional reduction scheme this being the most
appropriate to study an anomaly-free supersymmetric theory.  

At two-loops infinities have been
calculated in the amplitudes for $ D\geq 7$~\cite{BDDPR} 
(including
the $D=11$ case. The eleven dimensional counterterm was subsequently
evaluated in ref.~\cite{DS}). The expectation is that this theory will
be perturbatively infinite in four dimensions~\cite{NeightCounter}.

In this paper we shall examine features of the other interesting
extended supergravity.  
This is the $D=10$, $N=1$ theory and its
dimensional reduction
descendants which include the $D=4$, $N=4$ supergravity
theory (with a specific matter content). 
This supergravity is  the gravitational sector of the low
energy limit of both type~I and heterotic string theories
\cite{Heterotic}. 
For the one-loop amplitudes we may consider the $D=10\, ,\, N=1$ 
supergravity  together with arbitrary matter multiplets, 
although having less supersymmetry than the
$D=11,N=1$ theory, it is not clear which is the most fundamental. 
A web of dualities relates the various string theories and if one introduces
for instance the appropriate gauge group
SO(32) in the type I matter sector, this choice cancels the gravitational and 
gauge anomalies.  
When compactified to $D \leq 10$ type I and type II theories give rise to very
special  and similar symmetry groups.  
In fact it can be argued that these duality symmetries that have gained 
importance  
together with the realisation of the overwhelming necessity of
 non-perturbative effects are not that different between type I and type II
supergravities. The tantalizing $E_{10}$ is closely related to the hyperbolic
Kac-Moody algebra obtained from the extrapolation of the Chamseddine-Sen 
sequence \cite{J85}  namely the hyperbolic extension of $D_8$. In fact 
the symmetry groups of type I are 
fixed point sets under an involution of those 
of type II for all dimensions (at least three); 
this was mentioned in \cite{J9805} and relies on
Kac's description of the automorphisms of simple Lie algebras
which can be found for instance in \cite{Helg} . 

Clearly, the extended supergravities can
only be UV finite if they possess symmetries which we do not fully
understand yet, the implications of which will be at the non-perturbative 
level, or at least will require novel renormalisation techniques. 
Consequently, if these symmetries are operative at all, they should be 
approximately as powerful for type I and II theories. And one could 
infer from this that the structure of divergences should be somehow
related in both 
theories which are  two dual perturbative expansions. One should keep
in mind though that the relation is nonperturbative. 
 
The explicit (perturbative) calculation we shall perform is
four graviton scattering at one-loop for $D \geq 4$. This will
enable us to evaluate $R^4$ counterterms.
For the one-loop amplitudes we may consider the $D=10,N=1$ 
supergravity multiplet with arbitrary matter, we shall call it type I slightly 
abusively.  Although the theory will
only be free from gravitational anomalies in $D=10$ for special gauge groups, 
for $D=8\, , \, 10$ the gravitational anomalies only manifest
themselves in the
five or six graviton amplitude. Furthermore the contribution is parity
odd and finite whereas we shall consider divergences.

Before calculating the counterterms, we shall calculate the entire
amplitudes (including their finite part) and at first
 for a particular choice of
external helicities - namely in the case where the external
polarisation vectors are forced to be four dimensional. With this
simplification it is possible to present the amplitude in a very
elegant form as the sum over a few simple integral functions. From
these amplitudes we can easily see the presence of ultra-violet
infinities in $D=8\, , \, 10$.

Unlike the situation in four dimensions, there are onshell  
independent $R^4$ tensors in dimension  $D \geq 8$, so 
 in $D=8$  the counterterms are not determined
by  single coefficients and it requires a full computation of the
amplitude to fix their form. We find a beautiful
factorisation of the infinite counterterms in the amplitude
into a product of left times right
kinematic factors in a manner very reminiscent of the relationships
between gravity and Yang-Mills presented in ref.~\cite{BDDPR}. 
The
factorisation can be understood from a string theory viewpoint
but remains  more obscure for  field theorists.  Offshell this
factorisation is also not obvious at all from examining the field theory
counterterms although the counterterms can be manipulated to reflect it.


\section{ $D$ dimensional amplitudes with helicities in four dimensions}

As a simple case, we  first study amplitudes where the helicity
of the external gravitons are restricted to lie in the
four-dimensional space defined by the momenta. In this
situation we can calculate the entire amplitude in a fairly
simple form by breaking the amplitude into its helicity
components.  Even with the external helicities specified there are
different contributions depending upon which supermultiplet is
circulating in the internal loop. One may label the possible loop
contributions according to the 
circulating four dimensional supermultiplets. 
The three contributions we shall distinguish
 are that from a $N=8$ multiplet, that from
a $N=6$ matter multiplet and finally that from a $N=4$ matter
multiplet.  In terms of these contributions the
(pure) $N=4$ supergravity one-loop amplitude is
$$
M^{N=4}= M^{N=8}- 4M^{N=6,matter}   + 2 M^{N=4,matter}
\equn
$$
Also, of more interest to us is the ``$N=4^*$'' one-loop amplitude for the
dimensional reduction of $N=1,D=10$ supergravity to four dimensions which is
$$
\eqalign{
M^{N=4*} &= M^{N=8}- 4M^{N=6,matter}   + 8 M^{N=4,matter}
\cr}
\equn 
$$
It is often the case that nonmaximally supersymmetric theories do not remain 
irreducible upon dimensional reduction. 
If we consider the $N=4*$ (or type I) supergravity coupled to $N=4$
super-Yang-Mills matter the corresponding amplitude 
is
$$
\eqalign{
M^{N=4*, G} &= M^{N=8}- 4M^{N=6,matter}   + (8+{\rm dim}G ) M^{N=4,matter}
\cr}
\equn 
$$
where ${\rm dim}G$ is the dimension of the gauge group.  
Let us call $g$ the combination $g=(8+{\rm dim}G )$.
We have chosen to organize our amplitudes conveniently as  
linear combinations of the three supermultiplet contributions.

 For $D=10$ 
the three independent supersymmetric contributions are for instance
the $N=8$ and  the $N=4^*$ contributions as well as the ``matter'' $N=4$ term.  
These have in turn reductions below 10 dimensions.
If the external polarisations are forced to be
four dimensional then  the external gravitons 
have $\pm $ helicity. 
There are then only three independent ``helicity
amplitudes'', $M(1^+,2^+,3^+,4^+)$, $M(1^-,2^+,3^+,4^+)$ and
$M(1^-,2^-,3^+,4^+)$. (We choose a convention where all particles are
considered outgoing.)  In any supersymmetric theory,
$$
M(1^+,2^+,3^+,4^+) =M(1^-,2^+,3^+,4^+) = 0
\equn
$$
and the only non-zero independent amplitude is $M(1^-,2^-,3^+,4^+)$.  
We have calculated our amplitudes using ``String-based rules''
\cite{Long,Review,BDS,DunNor}, (see appendix~A)
and verified the four-dimensional expressions using unitarity 
techniques.
The details are presented in appendix~B. 
The amplitudes are all regulated 
using dimensional reduction \cite{DRED}
with parameter $\eps=(D-D')/2$.

Calculating the $N=8$ amplitude gives
$$
M^{N=8}(1,2,3,4) 
= \kaps 
[ stu M^{tree} ]  \left(4\pi \right)^{-D/2} 
\times   \left( I^{D}_4(s,t) +I^{D}_4(s,u) +I^{D}_4(t,u) \right)
\equn\label{EQeight} 
$$ 
where $I^{D}_4(s,t)$ denotes the $D$-dimensional scalar box integral
function with ordering of legs $1234$, $s,t$ and $u$ are the usual Mandelstam 
variables. The $n$-point scalar amplitude with external legs $k_i$ 
defined by 
$$
{\cal I}^{D}_n
=\int { d^Dp  \over (2\pi)^D }
{ 1 \over p^2 (p-k_1)^2 \cdots (p-\sum_{i=1}^{n-1} k_i)^2 }
\equn
$$
is related to the function $I_n^D$ by
$$
{\cal I}^{D}_n = i { (-1)^n \over (4\pi)^{D/2} }  I_n^D
\equn
$$

In general this is a function of the kinematic variables 
$(\sum_{a}^{b} k_i)^2$. Often we will express these variables which 
indicate the
ordering of legs. For example $I^{D}_4(s,t)$ has ordering of legs $1234$, 
$I^{D}_4(s,u)$ has ordering of legs $1243$ etc... 
The $N=8$  amplitude 
was originally evaluated using the low energy limit of string theory
by Green, Schwarz and Brink~\cite{GSB}.  Unlike the following two amplitudes
it is valid for all external polarisations.

The remaining two amplitudes are more complicated and have the form
$$
\eqalign{
M^{N=6,matter}(1^-,2^-,3^+,4^+) 
&=- \kaps [  stu  M^{tree} ] \times {(4\pi)^{-D/2}} \times  \biggl(
- \frac{1}{s} I_{4}^{D+2}(t,u)
\cr 
& \null \hskip 1.5 truecm 
+ \frac{(D-4)}{2s}  \left(  I^{D+2}_{4}(s,t) + I^{D+2}_{4}(s,u)
- I^{D+2}_{4}(t,u) 
\right) \biggr)\cr}
\equn\label{EQmattersix} 
$$ 

$$
\eqalign{
M^{N=4,matter}(1^-,2^-,3^+,4^+)
&=\kaps [ stu  M^{tree} ] \times  ({4\pi})^{-D/2} \times \biggl(
-{1 \over 2 s^2 } \left( I_2^D(t) +I_2^D(u) \right)
\cr
\null & \hskip -2.0 truecm
+{(D(D+2)) \over 4s^2  }I^{D+4}_4(u,t)+
{ (2-D)(4-D)  \over 4 s^2} I^{D+4}_4(s,t)+{ (2-D)(4-D)  \over 4 s^2} I^{D+4}_4(s,u)
  \biggr)
\cr}
\equn\label{EQmatterfour} 
$$ 
Note the appearance of
the $D$-dimensional scalar bubble integral,  
$I_2^D(t)$.

The tree amplitude for four-dimensional helicities  is 
$$
 M^{\rm tree}(1^-,2^-,3^+,4^+) \, =   i \; 
\kaps { st \over u} \left( \, 
{\spa1.2^4 \over \spa1.2\spa2.3\spa3.4\spa4.1} 
\right)^2\,.
\equn\label{OneLoopRelation}
$$
where we have expressed the amplitude using a ``spinor helicity''
representation of the polarisation tensors. Spinor helicity techniques
\cite{SpinorHelicity} were introduced for QCD calculation but by
splitting the polarisation tensor
$$
\epsilon^{\mu\nu} = \eps^{\mu} \bar{\epsilon}^{\nu}
\equn
$$
they can be applied to gravity calculations also 
\cite{BGK,BernGrant}.

These amplitudes are complete.  They contain a great deal of information but
in particular we  can simply extract their 
ultra-violet infinities. 
Since we are using dimensional
regularisation the one-loop integrals are only divergent in even
dimensions. For a four point amplitude the expected counterterm is of
the $R^4$ type  which can only appear for $D \geq 8$ at one-loop.  

For the $N=8$ amplitude the form of the amplitude is manifestly finite for
$D<8$ since scalar box integrals are only divergent for $D\geq 8$. 
Extracting the
divergences for $D=8$ and $D=10$ from the integrals yields
$$
\eqalign{
M^{N=8, D=8-2\eps}
&=\kaps { 1\over (4\pi )^4}\;
 [  stu M^{tree} ]\times   { 1 \over 2 \eps}
\cr
M^{N=8, D=10-2\eps}
&= 0 
\cr}
\equn
$$
The $D=10$ one-loop amplitude is onshell-convergent within
dimensional regularisation since the integrals give $\sim
(s+t+u)=0$.  A cut-off regularisation would give a non-zero result
proportional to 
$$
\Lambda^2 \times stu M^{tree} 
\equn
$$
which leads to the well known \cite{Tani,Tsey,GreenVH,RussoTsey} 
counterterm $\Lambda^2 t_8t_8 R^4$. 
($t_8$ will be defined later.) 
Within dimensional regularisation the
two-loop amplitude is the first divergence \cite{BDDPR}.

For the remaining two amplitudes we can evaluate the infinities in 
these expressions quite easily. The expressions are manifestly
convergent for $D < 6$ for the $N=6$ contribution and $D < 4$ for the
$N=4$ one.
Extracting the divergences we find
$$
\eqalign{
M^{N=6,matter,D=6-2\eps} ( 1^-,2^-,3^+ ,4^+ )
& =0 
\cr
M^{N=6,matter,D=8-2\eps} ( 1^-,2^-,3^+ ,4^+ )
&=\kaps { 1\over (4\pi )^4} [ stu  M^{tree}  ] \times   -{ 1 \over 24  \eps}
\cr
M^{N=6,matter,D=10-2\eps} ( 1^-,2^-,3^+ ,4^+ )
&=\kaps { 1\over (4\pi )^5} [ stu  M^{tree}  ] \times    -{ s \over 720  \eps}
\cr
M^{N=4,matter,D=6-2\eps} ( 1^-,2^-,3^+ ,4^+ )
&=0
\cr
M^{N=4,matter,D=8-2\eps} ( 1^-,2^-,3^+ ,4^+ )
&=
\kaps { 1\over (4\pi )^4} [   stu  M^{tree} ] \times    { 1 \over 180  \eps  } 
\cr
M^{N=4,matter,D=10-2\eps} ( 1^-,2^-,3^+ ,4^+ )
&=
\kaps { 1\over (4\pi )^5} [   stu  M^{tree} ] \times   {  s  \over 3360  \eps } 
\cr}
\equn
$$
In  both cases the $D=6$ counterterm 
vanishes. In $D=6$ the expected counterterm is $R^3$. 
However there is no supersymmetrisable $R^3$ counterterm and 
amplitudes contain no infinities. This is exactly the reason
why the two-loop
infinity vanishes in $D=4$ supergravity.

Finally, recombining these poles to give the 
full physical amplitude with four dimensional helicities
within type~I supergravity 
we find
$$
\eqalign{
M^{N=4*} ( 1^-,2^-,3^+& ,4^+ )
=
\kaps 
[ stu M^{\rm tree} ]  \times (4\pi)^{-D/2} 
\biggl(( I^{D}_4(s,t) +I^{D}_4(s,u) +I^{D}_4(t,u) )
\cr
&-{4  \over s} I_4^{D+2}(t,u)
- {2(4-D)  \over  s} \Bigl[   I^{D+2}_{4}(s,t) + I^{D+2}_{4}(s,u)
- I^{D+2}_{4}(t,u) ) \Bigr]
\cr
&+
{ 2(2-D)(4-D)  \over s^2} I^{D+4}_4(s,t) 
+{ 2(2-D)(4-D)  \over s^2} I^{D+4}_4(s,u)
 +{2D(D+2 )) \over  s^2 }I^{D+4}_4(u,t)
\cr
&-{4 \over  s^2} ( I_2^D(t) +I_2^D(u) )
\biggr)
\cr}
\equn
$$
with divergences
$$
\eqalign{
M^{N=4*,D=6} ( 1^-,2^-,3^+ ,4^+ )
&= 0 
\cr
M^{N=4*,D=8} ( 1^-,2^-,3^+ ,4^+ )
&= 
\kaps { 1\over (4\pi )^4}
[ stu M^{\rm tree} ] \times    { 32 \over 45  \eps} 
\cr
M^{N=4*,D=10} ( 1^-,2^-,3^+ ,4^+ )
&= 
\kaps { 1\over (4\pi )^5}
[ stu M^{\rm tree} ] \times   { s\over  126  \eps}
\cr}
\equn
$$

If these are coupled to super-Yang-Mills the divergences become
$$
\eqalign{
M^{N=4*,D=8} ( 1^-,2^-,3^+& ,4^+ )
= 
\kaps { 1\over (4\pi )^4}
[ stu M^{\rm tree} ] \times    { 120 + g \over 180  \eps} 
\cr
M^{N=4*,D=10} ( 1^-,2^-,3^+& ,4^+ )
= \kaps { 1\over (4\pi )^5}
[ stu M^{\rm tree} ] \times   { (3g +56)s\over  10080  \eps}
\cr}\equn
$$
where $g=(8+{\rm dim}G )$.

Although these amplitudes are indicative of the divergences present
they are not sufficient to determine completely
 the structure of the counterterms in higher dimensions.
For example in $D=8$ there are $7$ independent $R^4$ tensors
(onshell) compared to $2$ in $D=4$. When restricting the external
helicities 
and momenta to
$D=4$ naturally we loose most of the information. So to determine the
exact counterterms we must calculate with arbitrary external
helicity. This we shall do for $D=8\, ,\, 10$ in the following sections. 
We can also immediately see one interesting fact. For $N=4$ the $D=10$
divergence does not cancel - unlike in the $N=8$ case. This implies that
the counterterms must have a different $R^4$ structure. We shall
explore this further also.

Let us 
conclude the present section with the remark that the strange
coefficients 
appearing in 
the above formulas will actually be more reasonable looking in the higher 
dimensional amplitudes and result 
from  the collapse of several invariants onto the same expression in lower 
dimension.

\section{$D=8$ Counterterms}

In this section we move on from the complete amplitudes and focus upon
their infinity structure. We will relax our restriction
to four dimensional helicities and obtain the infinity in the
amplitudes in $D=8 \, , \, 10$ for arbitrary external polarisations.  These
amplitudes were calculated using the String-based method of
ref.~\cite{Long,BDS,DunNor}.  In this section we examine the $D=8$
infinity structure. The dimensional reduction of $D=10\, ,\, N=1$ to $D=8$
is  $D=8\, ,\, N=1$ supergravity~\cite{DeightSugra} plus matter.

In $D=8$ the potential one-loop counterterm is an $R^4$ tensor. This
is analogous to the situation in $D=4$ at three loops where a
potential $R^4$ term exists for $N=1$ supergravity~\cite{NeightCounter}. 
However, four
dimensions is rather special because many of the potentially
inequivalent $R^4$ 
tensors become equivalent at low dimensions. In
fact, and as alluded to above, from the potential seven tensors onshell 
(actually six remain after integration by parts) only two 
are inequivalent in four
dimensions. Of these only one is compatible with supersymmetry - this
is the well known Bel-Robinson tensor~\cite{BelRobinson}. However for 
$D \geq 8$ 
all seven tensors are inequivalent and the structure of $R^4$ tensors is much 
richer. The supersymmetrisability has been discussed in \cite{dR}.

Forgetting about supersymmetry we know from
\cite{Fulling} that a general $R^4$ tensor in $D=8$ is
$$
a_1 T_1 + a_2 T_2 +a_3 T_3 +a_4 T_4 + a_5 T_5 +a_6 T_6 +a_7 T_7
\equn\label{GeneralCounter}
$$
where
$$
\eqalign{
T_1 =& ( R_{p,q,r,s}R_{p,q,r,s})^2 
\cr
T_2 =& ( R_{p,q,r,s}R_{p,q,r,t})( R_{p',q',r',s}R_{p',q',r',t})
\cr
T_3 =& R_{p,q,r,s}R_{p,q,t,u}R_{t,u,v,w}R_{r,s,v,w}
\cr
T_4 =& R_{p,q,r,s}R_{p,q,t,u}R_{r,t,v,w}R_{s,u,v,w}
\cr
T_5 =& R_{p,q,r,s}R_{p,q,t,u}R_{r,v,t,w}R_{s,v,u,w}
\cr
T_6 =& R_{p,q,r,s}R_{p,t,r,u}R_{t,v,u,w}R_{q,v,s,w}
\cr
T_7=& R_{p,q,r,s}R_{p,t,r,u}R_{t,v,q,w}R_{u,v,s,w}
\cr}
\equn$$
These are onshell the independent tensors (actually the Riemann tensor means 
the Weyl tensor here) and the combination
$$
-{ T_1 \over 16}
+{T_2}
-{T_3  \over  8}
-T_4+2T_5-T_6 +2T_7
\equn$$
vanishes (or rather is a total divergence)
 being proportional to the Euler form.
$$
E \sim 
\eps_{a_1a_2a_3a_4a_5a_6a_7a_8}
\eps^{b_1b_2b_3b_4b_5b_6b_7b_8}
R^{a_1a_2}{}_{b_1b_2}
R^{a_3a_4}{}_{b_3b_4}
R^{a_5a_6}{}_{b_5b_6}
R^{a_7a_8}{}_{b_7b_8}
\equn$$

In order to calculate the appropriate $N=8$ counterterm we evaluate the 
(on-shell) amplitude and we find it factorises in the following way: 
$$
M^{N=8,D=8}
={1\over \epsilon} \times 
\kapss { i \over (4\pi )^4}\; { 1 \over 2}  K_1 \times K_1
\equn$$
where
$$
K_1 =  tu (\eps_1 \cdot \eps_2) (\eps_3 \cdot \eps_4)
+2 (\eps_1 \cdot \eps_2)  
\biggl( t(\eps_3\cdot k_1 \eps_4\cdot k_2)
+u(\eps_3  \cdot k_2   \eps_4\cdot k_1 )
\biggr) +\cdots
\equn$$
where $\cdots$ denotes symmetrisation over the four indices $1234$.
The factorisation is easily understood if one regards $N=8$
supergravity as the low energy limit of string theory, however it is
much more obscure from a field theory viewpoint. In fact the tensor
$K_1$ appears in string tree and loop amplitudes. (See
ref.~\cite{GSW} eqs. (7.4.42) and (9.A.19) ). 
Its appearance in many diverse 
calculations is presumably due to the 
uniqueness of a tensor compatible with maximal supersymmetry.

The counterterm necessary to cancel this infinity is,
\def\NORMFACT#1{ {1 \over \eps } \left( {\kappa \over 2} \right)^{4} 
{ i \over (4\pi)^{#1} }}
$$
-\NORMFACT{4} C^{N=8,D=8}
\equn
$$
where
$$
C^{N=8,D=8} =
a_2 \left[
-{ T_1 \over 16}
+{T_2}
-{T_3  \over  8}
-T_4+2T_5-T_6 +2T_7
 \right]
-{ 1 \over 4 } 
 \left[ T_4
-4T_7 \right]
\equn
$$ 
There is one arbitrary coefficient $a_2$ since the Euler term vanishes
onshell. Choosing  $a_2 =-1/4$ we have
$$
C^{N=8,D=8} = - {1 \over 4 } 
\left[
-{ T_1 \over 16}
+{T_2}
-{T_3  \over  8}
-0.T_4+2T_5-T_6 -2T_7
 \right]
\equn$$
This combination is precisely, the tensor combination
$$
{ 1 \over 128.6} t_8t_8R^4
\equn$$
which appears in the derivative expansion of the $M$-theory 
effective action ( \cite{GSW,GreenVH} ). The tensor $t_8$ is defined in 
\cite{GSW}, here we drop its antisymmetric part. 
The reason why $t_8$ appears in trees as well as loops is connected 
to the form of the vertex operators and to triality.

Calculating with the (formal) $N=6\, ,\, matter$ counterterm we find the
infinity has the same tensor structure and is
$$
M^{N=6,D=8} = - { 1 \over 12 } 
M^{N=8,D=8}
\equn$$

Finally, and more interestingly, we consider the infinity arising from the
$N=4$ {\it matter} multiplet. This also factorises into the form, 
$$
M^{N=4,D=8} 
=   { 1 \over \eps}  \times  \kapss { i \over (4\pi )^4} 
\times { 1 \over 720} 
K_1 \times K_2
\equn$$
where
$$
\eqalign{
K_2
=&
-\eps_1\cdot \eps_2 \eps_3 \cdot \eps_4 
(3t^2+5tu+3u^2)
-\eps_1\cdot \eps_3 \eps_2 \cdot \eps_4 
(3s^2+5st+3t^2)
-\eps_1\cdot \eps_4 \eps_2 \cdot \eps_3 
(3s^2+5su+3u^2)
\cr
&
+2  \eps_1\cdot \eps_2 \Bigl(
3s \eps_3 \cdot k_4 \eps_4 \cdot k_3 
+t \eps_3 \cdot k_1 \eps_4 \cdot k_2 
+u \eps_3 \cdot k_2 \eps_4 \cdot k_1  \Bigr) +\cdots 
\cr
&-12(
k_2\cdot \eps_1  
k_1 \cdot\eps_2  
k_4\cdot\eps_3  
k_3\cdot\eps_4 
+ 
k _3\cdot\eps_1  k_4\cdot\eps_2  k_1\cdot\eps_3  k_2\cdot\eps_4 
+ k_4\cdot\eps_1  k_3\cdot\eps_2  k_2\cdot\eps_3  k_1\cdot\eps_4 ) \cr}
\equn$$
We have organised $K_2$ according to the number of $\eps_i\cdot
\eps_j$.  The $\cdots$ denotes symmetrising the terms with a single
$\eps_i\cdot \eps_j$.  The counterterms necessary to cancel this are
$$
- \NORMFACT{4} 
C^{N=4,D=8}
\equn
$$
where
$$
C^{N=4,D=8}
= 
{1 \over 11520} 
\left( -3 T_1 +24 T_2 -6 T_3 
+{4 T_4 } 
+0.T_5 +0.T_6 +32 T_7 
\right)
\equn
$$

We can relate this also to specific tensors contracted against $R^4$.
The tensor $t_8$ can be split into two pieces  $t_{(12)}$ and $t_{(48)}$
each 
having the same symmetry properties as $t_8$. The tensors 
$t_{(12)}$ and $t_{(48)}$
contain 12 and 48 quartic monomials in the $\delta$'s respectively.
They are the only two tensors which have the same symmetry properties of 
$t_8$ itself in eight dimensions \cite{GSW}. 
Specifically
$$
t_8\,=\,{ 1 \over 2 } \Bigl(t_{(12)}+t_{(48)} \Bigr) 
\equn
$$
where 
\def\sym4#1#2#3#4{(\delta^{#1#3}\delta^{#2#4}-\delta^{#1#4}\delta^{#2#3}) }
$$
\eqalign{
t_{(12)}^{ijklmnpq}=&- 
\biggl( 
(\delta^{ik}\delta^{jl}-\delta^{il}\delta^{jk})
(\delta^{mp}\delta^{nq}-\delta^{mq}\delta^{np})
+\sym4{k}{l}{m}{n}\sym4{p}{q}{i}{j}
\cr
& \null\hskip 2.0 truecm 
+\sym4{i}{j}{m}{n}\sym4{k}{l}{p}{q}
\biggr)
\cr 
t_{(48)}^{ijklmnpq}=&
\biggl( \delta^{jk}\delta^{lm}\delta^{np}\delta^{qi}
+\delta^{jm}\delta^{nk}\delta^{lp}\delta^{qi}
+\delta^{jm}\delta^{np}\delta^{qk}\delta^{li}
+[i\leftrightarrow j]+[k\leftrightarrow l]+[m\leftrightarrow n] 
\biggr)\cr
}\equn$$
where $[i\leftrightarrow j]$ denotes antisymmetrisation with respect to
$i$ and $j$.
From these tensors we can define
$$
\eqalign{
A\,=&\,\frac{1}{4}t_{(12)} t_{(12)}\cdot R^4 \cr 
B\,=&\,\frac{1}{4} t_{(12)} t_{(48)}\cdot R^4 \cr
C\,=&\,\frac{1}{4}t_{(48)} t_{(48)}\cdot R^4 \cr 
}
\equn$$
where the $\cdot$ denotes the usual contraction of the upper and lower
eight indices.

\noindent We can also express these tensor contractions as traces
\cite{Tsey}
$$
t_{8} t_{(12)}\cdot R^4= 48t_8 \Tr ( R^4 ) 
\;\;\; ; \; \;\;
t_{8}t_{(48)}\cdot R^4= -12t_8  \Tr( R^2)\Tr( R^2)
\equn
$$

\noindent
In terms of these combinations the $N=8$ counterterm of the type
 $t_8t_8 R^4$ is just
$$
C^{N=8,D=8} = {1 \over 768} \Bigl( A+2B+C \Bigr) 
\equn$$

\noindent
The $N=4$ counterterm can also be expressed in terms of $A\, ,\, B$ and $C$ 
and is
$$
C^{N=4,D=8} = \ {} {1 \over 276480}  \Bigl(
-5A -4B +C 
\Bigr) 
\equn$$

\noindent
This second tensor structure has also played a role in the low energy limit
of string theory, although in this case that of heterotic or type~I
string theory.  It has appeared in the ten dimensional effective
action as seen by \cite{Kikuchi1986,Cai:1987sa,Ellis:1988dc} both as a
string tree and one-loop correction.  Again the string formulation
of tree amplitudes and thus of the appropriate invariant tensors
gives an understanding of the factorisation of the amplitude which does not
have a simple interpretation in field theory.
One checks easily that $5A+4B-C$ contains a $t_8$ factor.
 The counterterm is also 
supersymmetrisable as shown in \cite{dR} where the most general 
lagrangians of the form $R+R^4$ were considered (without vector multiplets).

\noindent
Considering $N=4^*$ supergravity coupled to matter we find
$$
C^{N=4*,D=8}
={( g + 480 ) \over 2^{11}.3^3.5 }\Bigl( A+2B+C \Bigr) 
-{ 6g \over 2^{11}.3^3.5 }\Bigl( A+B \Bigr) 
\equn$$
where $g=(8+{\rm dim}G )$.

\section{$D=10$ Counterterms}

Ten dimensions is, of course, the natural home of 
both $N=4^*$ supergravity
and of superstring theory.  However,
within dimensional regularisation it is the $D=8$ counterterms that 
match easily tensors which appear in string theory. In
dimensional regularisation possible counterterms in $D$ dimensions at
$L$ loops are
$$
 \partial^n R^m
\equn$$
where $n+2m = (D-2)L+2$. 
With a cut-off regulator, string theory being a physical regulator,
 the equivalent terms are
$$
 \Lambda^n R^m
\equn$$
For $D=10$ our counterterms are thus of the form $\partial^2 R^4$. Indices
have been suppressed in this expression and the full form is
$$
T^{\alpha_1\alpha_2\mu_1\nu_1\rho_1\sigma_1\mu_2\nu_2\rho_2\sigma_2\mu_3\nu_3\rho_3\sigma_3\mu_4\nu_4\rho_4\sigma_4}
\partial_{\alpha_1} 
R_{\mu_1\nu_1\rho_1\sigma_1}
\partial_{\alpha_2}R_{\mu_2\nu_2\rho_2\sigma_2}
R_{\mu_3\nu_3\rho_3\sigma_3}
R_{\mu_4\nu_4\rho_4\sigma_4}
\equn
$$
Unless the tensor splits as follows
$$
T^{\alpha_1\alpha_2\mu_1\nu_1\rho_1\sigma_1\mu_2\nu_2\rho_2\sigma_2
\mu_3\nu_3\rho_3\sigma_3\mu_4\nu_4\rho_4\sigma_4}
\sim 
\delta^{\alpha_1\alpha_2}
T^{\mu_1\nu_1\rho_1\sigma_1\mu_2\nu_2\rho_2\sigma_2
\mu_3\nu_3\rho_3\sigma_3\mu_4\nu_4\rho_4\sigma_4}
\equn$$
it will not be expressible in terms of the set of tensors found in
$D=8$ and those arising in string theory. 
As a matter of fact one of the counterterms we find does not split
in this way.

Recall that in $D=10$ the ``physical'' combinations are
$$
\eqalign{
 &M^{N=8,D=10}
\cr
 &M^{N=4,D=10} 
\cr
 &M^{N=4,D=10*} = M^{N=8,D=10}
-4M^{N=6,D=10} 
+8 M^{N=4,D=10} 
\cr}
\equn$$
The results of calculating the infinities are firstly,
$$
M^{N=8,D=10}=0 \equn$$ as expected.  The infinities have two powers
more of momentum as compared to $D=8$. However we still find that all 
infinities factorise with one factor of $K_1$. The remaining factor
$L_i$ contains the extra two powers of momentum. Specifically we
calculate
$$
\eqalign{
M^{N=4,D=10} &=  {1\over \epsilon}\times \kapss 
{ -i \over (4\pi )^5}
 { 1 \over 60480  }
K_1 \times L_1
\cr
M^{N=6,D=10} &= {1\over \epsilon}\times \kapss 
{ -i \over (4\pi )^5} 
  { 1 \over 1440  }
K_1 \times L_2 
\cr}
\equn$$
where
\small
$$
\eqalign{ 
L_1 &=(\eps_1 \cdot \eps_2)(\eps_3 \cdot \eps_4)
s( 18u^2+41tu +18t^2 ) +\cdots
\cr
& \null \hskip -0.5 truecm
+2(\eps_1 \cdot \eps_2) 
\Bigl(-t^2(18\eps_3\cdot k_4 \eps_4\cdot k_3 +\eps_3\cdot k_1 \eps_4\cdot
k_2 )
-u^2(18\eps_3\cdot k_4 \eps_4\cdot k_3 +
\eps_3\cdot k_2 \eps_4\cdot k_1 )  
\cr
& \null \hskip 0.5 truecm
-tu(40\eps_3\cdot k_4 \eps_4\cdot k_3 +\eps_3\cdot k_1 \eps_4\cdot k_2 +\eps_3\cdot k_2 
\eps_4\cdot k_1 ) \Bigr) +\cdots
\cr
& \null \hskip -0.5 truecm
+4\Bigl(
(t\eps_1\cdot k_3 \eps_2\cdot k_1 \eps_3\cdot k_2 \eps_4\cdot k_1 
+5t\eps_1\cdot k_3 \eps_2\cdot k_1 \eps_3\cdot k_2 \eps_4\cdot k_2 
+6t\eps_1\cdot k_3 \eps_2\cdot k_3 \eps_3\cdot k_1 \eps_4\cdot k_1 
+t\eps_1\cdot k_3 \eps_2\cdot k_3 \eps_3\cdot k_1 \eps_4\cdot k_2 
\cr
& \null \hskip -0.5 truecm
-17t\eps_1\cdot k_3 \eps_2\cdot k_3 \eps_3\cdot k_2 \eps_4\cdot k_1 
+6t\eps_1\cdot k_3 \eps_2\cdot k_3 \eps_3\cdot k_2 \eps_4\cdot k_2 
-18t\eps_1\cdot k_2 \eps_2\cdot k_1 \eps_3\cdot k_2 \eps_4\cdot k_1 
-18t\eps_1\cdot k_2 \eps_2\cdot k_1 \eps_3\cdot k_2 \eps_4\cdot k_2 
\cr
& \null \hskip -0.5 truecm
+t\eps_1\cdot k_2 \eps_2\cdot k_3 \eps_3\cdot k_1 \eps_4\cdot k_1 
-4t\eps_1\cdot k_2 \eps_2\cdot k_3 \eps_3\cdot k_1 \eps_4\cdot k_2 
-18t\eps_1\cdot k_2 \eps_2\cdot k_3 \eps_3\cdot k_2 \eps_4\cdot k_1 
-23t\eps_1\cdot k_2 \eps_2\cdot k_1 \eps_3\cdot k_1 \eps_4\cdot k_2 
\cr
& \null \hskip -0.5 truecm
-17t\eps_1\cdot k_2 \eps_2\cdot k_1 \eps_3\cdot k_1 \eps_4\cdot k_1 
+5t\eps_1\cdot k_3 \eps_2\cdot k_1 \eps_3\cdot k_1 \eps_4\cdot k_2 
+6t\eps_1\cdot k_3 \eps_2\cdot k_1 \eps_3\cdot k_1 \eps_4\cdot k_1 
-18u\eps_1\cdot k_2 \eps_2\cdot k_1 \eps_3\cdot k_1 \eps_4\cdot k_1 
\cr
& \null \hskip -0.5 truecm
-18u\eps_1\cdot k_3 \eps_2\cdot k_1 \eps_3\cdot k_1 \eps_4\cdot k_2 
+6u\eps_1\cdot k_3 \eps_2\cdot k_3 \eps_3\cdot k_1 \eps_4\cdot k_1 
-23u\eps_1\cdot k_2 \eps_2\cdot k_1 \eps_3\cdot k_2 \eps_4\cdot k_1 
-17u\eps_1\cdot k_2 \eps_2\cdot k_1 \eps_3\cdot k_2 \eps_4\cdot k_2 
\cr
& \null \hskip -0.5 truecm
+5u\eps_1\cdot k_2 \eps_2\cdot k_3 \eps_3\cdot k_1 \eps_4\cdot k_1 
+u\eps_1\cdot k_2 \eps_2\cdot k_3 \eps_3\cdot k_1 \eps_4\cdot k_2 
+5u\eps_1\cdot k_2 \eps_2\cdot k_3 \eps_3\cdot k_2 \eps_4\cdot k_1 
+6u\eps_1\cdot k_2 \eps_2\cdot k_3 \eps_3\cdot k_2 \eps_4\cdot k_2 
\cr
& \null \hskip -0.5 truecm
-17u\eps_1\cdot k_3 \eps_2\cdot k_3 \eps_3\cdot k_1 \eps_4\cdot k_2 
+u\eps_1\cdot k_3 \eps_2\cdot k_3 \eps_3\cdot k_2 \eps_4\cdot k_1 
+6u\eps_1\cdot k_3 \eps_2\cdot k_3 \eps_3\cdot k_2 \eps_4\cdot k_2 
\cr
& \null \hskip -0.5 truecm
-4u\eps_1\cdot k_3 \eps_2\cdot k_1 \eps_3\cdot k_2 \eps_4\cdot k_1 
+u\eps_1\cdot k_3 \eps_2\cdot k_1 \eps_3\cdot k_2 \eps_4\cdot k_2 
-18u\eps_1\cdot k_2 \eps_2\cdot k_1 \eps_3\cdot k_1 \eps_4\cdot k_2 \Bigr)
\cr}
\equn$$
and
$$
\eqalign{
L_2 =& 
-(\eps_1 \cdot \eps_2)  (\eps_3 \cdot \eps_4)
 s(2u^2 +3tu +2t^2 ) +\cdots
\cr
\null & \hskip -0.5 truecm
+2( \eps_1 \cdot \eps_2)
\Bigl( t^2(2\eps_4\cdot k_3 \eps_3\cdot k_4 +\eps_3\cdot k_1 \eps_4\cdot k_2 )
+u^2(2\eps_34\cdot k_4 \eps_4\cdot k_3 +\eps_3\cdot k_2 \eps_4\cdot k_1 )
\cr
\null &
+tu(2\eps_3\cdot k_4 \eps_4\cdot k_3 +\eps_3\cdot k_1 \eps_4\cdot k_2 
+\eps_3\cdot k_2 \eps_4\cdot k_1 \Bigr) \cdots \cr
\null &\hskip -0.7 truecm
-4\Bigl(
-u\eps_1\cdot k_2 \eps_2\cdot k_3 \eps_3\cdot k_2 \eps_4\cdot k_1 
-u\eps_1\cdot k_3 \eps_2\cdot k_3 \eps_3\cdot k_1 \eps_4\cdot k_2 
+u\eps_1\cdot k_3 \eps_2\cdot k_3 \eps_3\cdot k_2 \eps_4\cdot k_1 
\cr
\null & \hskip -0.5 truecm
-u\eps_1\cdot k_2 \eps_2\cdot k_1 \eps_3\cdot k_2 \eps_4\cdot k_2 
-u\eps_1\cdot k_2 \eps_2\cdot k_3 \eps_3\cdot k_1 \eps_4\cdot k_1 
-2u\eps_1\cdot k_2 \eps_2\cdot k_1 \eps_3\cdot k_1 \eps_4\cdot k_1 
-2u\eps_1\cdot k_3 \eps_2\cdot k_1 \eps_3\cdot k_1 \eps_4\cdot k_2 
\cr
\null & \hskip -0.5 truecm 
+u\eps_1\cdot k_3 \eps_2\cdot k_1 \eps_3\cdot k_2 \eps_4\cdot k_2 
-u\eps_1\cdot k_2 \eps_2\cdot k_1 \eps_3\cdot k_2 \eps_4\cdot k_1 
+u\eps_1\cdot k_2 \eps_2\cdot k_3 \eps_3\cdot k_1 \eps_4\cdot k_2 
\cr
\null &  \hskip -0.5 truecm 
+2u\eps_1\cdot k_3 \eps_2\cdot k_1 \eps_3\cdot k_2 \eps_4\cdot k_1 
-2u\eps_1\cdot k_2 \eps_2\cdot k_1 \eps_3\cdot k_1 \eps_4\cdot k_2 
-2t\eps_1\cdot k_2 \eps_2\cdot k_1 \eps_3\cdot k_2 \eps_4\cdot k_1 
\cr
\null & \hskip -0.5 truecm 
-2t\eps_1\cdot k_2 \eps_2\cdot k_1 \eps_3\cdot k_2 \eps_4\cdot k_2 
+t\eps_1\cdot k_2 \eps_2\cdot k_3 \eps_3\cdot k_1 \eps_4\cdot k_1 
+2t\eps_1\cdot k_2 \eps_2\cdot k_3 \eps_3\cdot k_1 \eps_4\cdot k_2 
-2t\eps_1\cdot k_2 \eps_2\cdot k_3 \eps_3\cdot k_2 \eps_4\cdot k_1 
\cr
\null & \hskip -0.5 truecm 
-t\eps_1\cdot k_2 \eps_2\cdot k_1 \eps_3\cdot k_1 \eps_4\cdot k_2 
-t\eps_1\cdot k_2 \eps_2\cdot k_1 \eps_3\cdot k_1 \eps_4\cdot k_1 
-t\eps_1\cdot k_3 \eps_2\cdot k_1 \eps_3\cdot k_1 \eps_4\cdot k_2 
+t\eps_1\cdot k_3 \eps_2\cdot k_3 \eps_3\cdot k_1 \eps_4\cdot k_2 
\cr
\null & \hskip -0.5 truecm 
-t\eps_1\cdot k_3 \eps_2\cdot k_3 \eps_3\cdot k_2 \eps_4\cdot k_1 
+t\eps_1\cdot k_3 \eps_2\cdot k_1 \eps_3\cdot k_2 \eps_4\cdot k_1 
-t\eps_1\cdot k_3 \eps_2\cdot k_1 \eps_3\cdot k_2 \eps_4\cdot k_2  \Bigr) 
\cr}
\equn$$
\normalsize

The infinities can be canceled by specific combinations of $\partial^2
R^4$ counterterms.  Once more we find local Lorentz invariant
counterterms to cancel the infinities. As a working hypothesis we
assumed factorisability to make this tractable. The strategy is to
calculate arbitrary onshell scalars of the form $\partial^2 F^4$ times the
previous $t_8F^4$, to deduce from the resulting three parameter
expression for the amplitude the values of these coefficients 
and finally to replace the two polynomial solutions in the
$F$'s by the corresponding invariants in the fourth order in the
Riemann tensor. (Onshell equivalent to the Weyl tensor at this order.)
As is explained in detail in appendix~C we
may choose to express the counterterms in
terms of the following set of tensors,
$$
\eqalign{
S_{1} =& (\partial_{\alpha} R_{p,q,r,s}\partial_{\alpha}R_{p,q,r,s}) 
( R_{p',q',r',s'}R_{p',q',r',s'})
\cr
S_{2} =& (\partial_{\alpha} R_{p,q,r,s}\partial_{\alpha}R_{p,q,r,t})
( R_{p',q',r',s}R_{p',q',r',t})
\cr
S_{3} =& ( \partial_{\alpha}R_{p,q,r,s}R_{p,q,r,t})
( \partial_{\alpha}
R_{p',q',r',s}R_{p',q',r',t})
\cr
S_{4} =& \partial_{\alpha} R_{p,q,r,s}\partial^{\alpha}R_{p,q,t,u}
R_{t,u,v,w}R_{r,s,v,w}
\cr
S_{5} =& 
\partial_{\alpha} R_{p,q,r,s} \partial^{\alpha} R_{p,q,t,u}
R_{r,t,v,w}R_{s,u,v,w}
\cr
S_{6} =&\partial_{\alpha} R_{p,q,r,s}\partial^{\alpha} R_{p,q,t,u}
R_{r,v,t,w}R_{s,v,u,w}
\cr
S_{7} =&\partial_{\alpha} R_{p,q,r,s}\partial^{\alpha}R_{p,t,r,u}
R_{t,v,u,w}R_{q,v,s,w}
\cr
S_{8} =&\partial_{\alpha} R_{p,q,r,s}\partial^{\alpha}R_{p,t,r,u}
R_{t,v,q,w}R_{u,v,s,w}
\cr
S_9 =& R_{m,b,c,d}R_{n,b,c,d}
\partial_{m} R_{e,f,g,h} \partial_n R_{e,f,g,h}
\cr
S_{10}=& \partial_p R_{a,b,l,m} \partial_q R_{a,b,r,s}
R_{q,d,l,r}R_{p,d,m,s}
\cr}
\equn$$
For $S_1$ to $S_8$ the derivatives are contracted with each other and
these $S_i$'s are related to derivatives acting upon the $T_i$'s of the
previous sections. Tensors $S_9$ and $S_{10}$ however
 have the derivatives contracted
into the Riemann tensors. Of course many tensors of the form $\partial^2 R^4$
vanish onshell since they produce amplitudes of the form
$$
\sim ( s+t +u ) \times { \rm tensor } =0
\equn$$

In terms of the $S_i$'s  the infinities are canceled by the counterterms
$$
- \NORMFACT{5} C^{N=6,D=10} \hskip 1.0 truecm \hbox{\rm and}
\hskip 0.5 truecm
- \NORMFACT{5} C^{N=4,D=10}
\equn
$$
where
$$
\eqalign{
C^{N=6,D=10} 
&= { 1 \over 4. 720   }  \Bigl(  
 { S_1 } - 12S_2 - 4S_3  +2 {S_4}  
+ 0.S_5  - 8S_6  +16 S_7  +8S_8 \Bigr)
\cr
C^{N=4,D=10} 
& ={1  \over  4.6048   }\Bigl(-9S_{1} +76S_{2} -44 S_{3}-30S_{4}
\cr
\null& \null\hskip 3.5 truecm 
 +56S_{5} -88S_{6} -16 S_{7} +88S_{8}-24S_{9}+95S_{10}
\Bigr)
\cr}
\equn$$

It is far from obvious that such counterterms lead to infinities which 
factorise, however we can manipulate them to do so.
In fact it is possible to express both tensors in the form
$$
t_{10} t_8 \partial^2 R^2
 \equiv
t_{10}^{a_1a_2a_3a_4a_5a_6a_7a_8a_9a_{10}}
t_8{}_{b_1b_2b_3b_4b_5b_6b_7b_8} 
\partial_{a_1} R_{a_2a_3}{}^{b_1b_2}
 \partial_{a_4} R_{a_5a_6}{}^{b_3b_4}
R_{a_7a_8}{}^{b_5b_6}
R_{a_9a_{10}}{}^{b_7b_8}
\equn$$
where we have chosen to contract the derivatives into the $t_{10}$ tensor. 

The specific tensors are
$$
\eqalign{ 
t_{10}^{N=4,a_1a_2a_3a_4a_5a_6a_7a_8a_9a_{10}}
=&  10 \delta^{ a_1a_4}\delta^{a_2a_5}
\delta^{a_3a_6}\delta^{a_7a_9}\delta^{a_8a_{10}}
+4 \delta^{a_1a_4}\delta^{a_2a_{10}}\delta^{a_3a_5}
\delta^{a_6a_7}\delta^{a_8a_9}
\cr
\null \hskip 4.0 truecm &
+4\delta^{a_1a_{10}}\delta^{a_2a_5}\delta^{a_3a_6}
\delta^{a_4a_7}\delta^{a_8a_9}
\cr
t_{10}^{N=6,a_1a_2a_3a_4a_5a_6a_7a_8a_9a_{10}}
=&  \delta^{ a_1a_4}\delta^{a_2a_5}
\delta^{a_3a_6}\delta^{a_7a_9}\delta^{a_8a_{10}}
-4  \delta^{a_1a_4}\delta^{a_2a_{10}}\delta^{a_3a_5}
\delta^{a_6a_7}\delta^{a_8a_9}
\cr}
\equn$$
Where possible one must antisymmetrise with respect to the 
pairs of indices $a_2\leftrightarrow a_3 $ etc and symmetrise with respect to
pairs of couples of indices $(a_2a_3) \leftrightarrow (b_1b_2)$. 
The tensor for $N=4^*$ and $N=4^*$ coupled to matter being the
appropriate linear combination of these.  Both tensors define
$\partial^2 R^4$ tensor which must be $D=10,N=1$ supersymmetrisable.
Presumably the vanishing of the $M^{N=8,D=10}$ is because no 
$D=10,N=2$ supersymmetrisable $\partial^2 R^4$ tensor exists.

The counterterms for $D=10,N=1$ supergravities are given by  linear
combinations of the various $M^{D=10}$ and in fact the infinities are
given by $K_1 \times \sum_{i} c_i L_i$ since the various infinities
factorise in this way. At the level of the counterterms, all
type I
supergravity divergences contain the factor $t_8$.

\section{Conclusions}

In this letter we have determined the one-loop ultra-violet behavior
of the dimensional reductions of $D=10\, ,\, N=1$ supergravity. We have
examined this by calculations involving physical on-shell (four-point)
amplitudes.  We have found a variety of results; some in complete
agreement with expectations but some not so obvious 
from the field theory viewpoint. 
As expected, the amplitudes are one-loop finite for $D < 8$. For $D
\geq 8$ there is a richer structure. In $D=8$ we find infinities which
correspond to $R^4$ counterterms and we have completely determined
the $R^4$ structure. In $D=4$ the structure of $R^4$ terms is fairly
simple; there are only two independent tensors of which one is
compatible with supersymmetry - the well known ``Bel-Robinson''
combination~\cite{BelRobinson}.   
For $D \geq 8$, there are seven
independent tensors.  For the $D=8 \, , \, N=1$ theory supersymmetry is less
restrictive than for $D=8\, , \, N=2$ and this 
allows a further $R^4$ counterterm, in fact 
we do find infinities belonging to a new
tensor structure.  

Within dimensional regularisation, $D=10$ counterterms to
a four-point infinity are of the form $\partial^2 R^4$. The vanishing
of the $D=10$, $N=2$ one-loop infinity is presumably due to the
non-existence of a supersymmetric combination of 
$\partial^2 R^4$ counterterms (which does not vanish
onshell.)  For $D=10,N=1$ we do however find a $\partial^2 R^4$
counterterm which is non-zero.

An interesting feature of the amplitudes (so far checked up to 1-loop level) 
is {\it factorisation}. For
the $D=8$ counterterms one finds that the infinities in the amplitude
factorise as $K_1\times K_2$ where the $K_a$ are combinations of the
external $\eps_i$ and $k_i$.  Similarly the $D=10$ counterterms factorise as
$K_1\times L_2$.  This factorisation is far from manifest when
examining the counterterms however. Regarding the amplitudes as
arising from the low energy limit of string theory, this factorisation
is the remnant of the string factorisation into left and right moving
amplitudes. 
However, it should be noted that the string factorisation
occurs {\it within} the loop momentum integral whereas the
factorisation of infinity occurs in the amplitude.

The factorisation 
is expected whenever the tree corrections generate the same invariants.
However the factorisation
of the amplitudes is unexpected from a field theory viewpoint and is
hinting towards an alternate description of gravity theories as a
product of two Yang-Mills theories - as suggested in
\cite{BDDPR,BernGrant}. Although such a formalism is natural in string
theory it might well exist in a purely field theoretic context.

The counterterm structures we find are related to terms appearing in
various places in string theory. This is not surprising since string
theory provides a physical regularisation of supergravity. Also
String/M theory shares many symmetries with supergravity theory and
hence counterterms/effective actions are subject to the same
constraints. In the present work we have calculated purely within a field 
theory context. The structures we find are presumably inherent to any
regulator of supergravity whether a string theory or a more conventional one.

Acknowledgements. We wish to thank Pascal Bain,
Zvi Bern, Paul Howe
and
David Kosower
for useful conversations.

\vfill\eject
\appendix

\section{Calculations}

We have calculated using the String-based techniques, originally
developed for Yang-Mills calculations
\cite{Long,BernDunbar,Review} and then applied to gravity calculations
\cite{BDS,DunNor}.  
Although originating in string theory the same formalism was
subsequently shown to arise from a ``world-line'' approach to field
theory~\cite{WorldLine}.  We refer the reader to these papers for the
details of the technique, here merely presenting the results for our
amplitudes.

The techniques provide a formalism for obtaining the Feynman parameter
integrals for on-shell amplitudes.
The initial step is to draw all $\phi^3$ diagrams,
excluding tadpoles.  There is no need to include diagrams with a
loop isolated on an external leg since these vanish when dimensional
regularisation is used.  The external legs of these diagrams should be
labeled, with diagrams containing all orderings included.  The inner
lines of trees attached to the loop are labeled according to the rule
that as one moves from the outer lines to the inner ones, one labels
the inner line with the same label as the most clockwise of the two
outer lines attached to it.  The contribution from each labeled
$n$-point $\phi^3$-like diagram with $n_{\ell}$ legs attached to the
loop is
$$
\eqalign{
{\cal D} =
i { (-\kappa)^n \over (4\pi)^{D/2} }
 \Gamma(n_\ell-D/2)
&\int_0^1 dx_{i_{n_\ell-1}} \int_0^{x_{i_{n_\ell-1}}} dx_{i_{n_\ell-2}} \cdots
\int_0^{x_{i_3}}  dx_{i_2} \int_0^{x_{i_2}} dx_{i_1} \cr
& \times
{K^{}_{\rm red}(x_{i_1},\dots,x_{i_{n_\ell}}) \over
\Bigl(\sum_{l<m}^{n_\ell} P_{i_l} P_{i_m} \x {i_m}{i_l}
(1-\x {i_m}{i_l})\Bigr)^{n_\ell -D/2}}  \cr}
\equn
$$
where the ordering of the loop parameter integrals corresponds to the
ordering of the $n_\ell$ lines attached to the loop,
$x_{ij} \equiv x_i - x_j$. The $x_{i_m}$
are related to ordinary Feynman parameters
by $x_{i_m} = \sum_{j=1}^m a_j$. $K_{\rm red}$ is the ``reduced
kinematic factor'', which the string-based rules efficiently yield
in a compact form.  The lines attached to the loop carry
momenta $P_i$ which will be off-shell if there is a tree attached to
that line.  

\def\Gbdb{\dot {\overline G}{}_B}
\def\Gbddb{\ddot {\overline G}{}_B}
\def\Gb{G_B}
\def\Gbd{\dot G_B}
\def\Gbdd{\ddot G_B}

One obtains $K_{\rm red}$ by applying substitution rules to an 
overall kinematic factor,   
$$
\eqalign{
{\cal K} &=
\int \prod_{i=1}^n dx_i d \bar x_i \prod_{i<j}^n
\exp\biggl[ k_i\cdot k_j G_B^{ij} \biggr]
\exp \biggl[ (k_i\cdot\pol_j - k_j\cdot\pol_i) \, \Gbd^{ij}
         - \pol_i\cdot\pol_j\, \Gbdd^{ij} \biggr] 
\cr
& \hskip 1 cm \times
\exp \biggl[ (k_i\cdot\bar\pol_j - k_j\cdot\bar\pol_i) \, \Gbdb^{ij}
         - \bar\pol_i\cdot\bar\pol_j\, \Gbddb^{ij} \biggr]
\exp \biggl[- ( \pol_i\cdot\bar\pol_j + \pol_j\cdot\bar\pol_i )
\,  H_B^{ij} \biggr]
\biggr|_{\rm multi-linear} 
\cr}
\equn\label{MasterKin}
$$ 
where the `multi-linear' indicates that only terms linear in all
$\pol_i$ and $\bar\pol_i$ are included.  The graviton polarization
tensor is reconstructed by taking 
$\pol_i^\mu\bar\pol_i^\nu\rightarrow \pol_i^{\mu\nu}$.  
From a string theory perspective
$G_B$ is the bosonic Green function on the string world sheet, $\Gbd$
and $\Gbdd$ are derivatives of this Green function with respect to
left-moving variables, and $\Gbdb$ and $\Gbddb$ are derivatives with
respect to right-moving ones.  (Since a closed string is periodic the
variables describing the string world sheet can be split into
``left-moving'' and ``right-moving''.)  The term $H_B^{ij}$ is the
derivative of the Green function with respect to one left mover and
one right mover variable.  The functions $G_B^{ij}$, $\Gbdd^{ij}$ and
$H_B^{ij}$ are to taken as symmetric in the $i$ and $j$ indices while
$\Gbd$ is antisymmetric.  Although the above expression contains much
information in string theory, when one takes the infinite string
tension limit it should merely be regarded as a function which
contains all the information necessary to generate $K_{red}$ for all
graphs.  The utility of the string based method partially lies in this
compact representation (which is valid for arbitrary numbers of
legs!).  The existence of an overall function which reduces to the
Feynman parameter polynomial for each diagram is one of the most
useful features of the String-based rules.

Slightly different substitutions rules are used
depending upon particle type. The technique is particularly simple for
supersymmetric theories where the cancellations between particle types
circulating are manifest.

In general in Gravity theories, in a one-loop $n$-point amplitude the
amplitude is a sum over diagrams with $n_\ell$ legs attached to the loop
where $n_\ell \leq n$. For these diagrams the integrand is a polynomial of
the Feynman parameters of degree $2n_\ell$.  The overlying kinematic
expression is of the form $\sum K_L \times K_R$.  The substitutions
act upon $K_L$ and $K_R$ independently - each being up to $n_\ell$ powers
of Feynman Parameters.  For a given particle circulating in the loop
the substitution is of the form
$$
K_{L/R}  \rightarrow \pm S + C_p
\equn
$$
where $S$ denotes a part common to all particle types and $C_p$ is a
``cycle'' contribution which depends upon particle type.  In general
$S$ is of degree $n_\ell$ whereas $C_P$ is typically $n_\ell-2$ or less.
Bosonic/fermionic particles have a $\pm S$ term. In a supersymmetric
theory we are thus guaranteed that the term $S \times S $ cancels and
the Feynman parameter integral has maximum degree $n_\ell-2$. For extended
supersymmetry there can be further cancellations between the $C_p$
terms and for a $n$-point amplitude the degree of the polynomial is
$2n_\ell-4$ for $N\geq 4$.  For supersymmetry with $N\geq 4$ the
cancellations imply that the only $\phi^3$ diagrams contributing to a
four point amplitude are the three box integrals. (This is not quite
obvious the cancellations occur on $K_L$ only~\cite{DunNor} so if
$n_\ell=3$ for example the $K_L$ becomes a polynomial of degree $3-4=-1$. In
other words it cancels completely. )

In ref.~\cite{DunNor} the low energy limit of string theory can be taken and the
loop contributions from the different particle types disentangled to obtain
the contributions due to a single graviton, Weyl fermion, scalar etc. Here we
will be reconstructing the $N=4^*$ amplitude again. A string theory consists of
two sectors - the Neveu-Schwarz sector (NS) and the Ramond sector (R). These 
sectors contribute in the low-energy limit
$$
\eqalign{
NS &\longrightarrow 8S +2C_V
\cr
R &\longrightarrow -8S-8C_F
\cr}
\equn$$
For a type~II superstring which has $N=8$ supergravity as its low energy limit
the contributions are of the form
$$
\left( NS +R \right) \times
\left( NS +R \right)
\equn$$
adds up to
$$
4[C_V-4C_F ; C_V -4C_F]
\equn$$
For a superstring
with $N=4^*$ as its low energy limit the contribution is
$$
\left( NS +R \right) \times
 NS^* 
\equn
$$
leading to a contribution
$$
2[C_V-4C_F ; S]
\equn
$$
We use four dimensional helicity to simplify the overall kinematic
expression whereas letting the particles circulate in any
dimension $4\leq D \leq 10$. This is very similar to the calculations of 
ref.~\cite{DunNor} although here we let $D$ vary from 4 to 10. 
The contributions we seek are for the $N=8$, $N=6$ and $N=4$ matter multiplets.

For the amplitudes $M(1^-,2^-,3^+,4^+)$ a choice of spinor helicity
basis simplifies the kinematic expression enormously.  Spinor Helicity
techniques utilize a representation of the polarisation vectors in terms
of spinor products,
$$
\epsilon^\pm_\mu(p,k) = \pm {\langle p \pm | \gamma_\mu | k \pm \rangle \over
\sqrt{2} \langle k \mp | p \pm \rangle}
\equn$$
where $p^{\mu}$ is the momentum of the external state and $k^{\mu}$ is a
``reference momentum'' satisfying $k^2=0$. (Different choices of
$k^{\mu}$ correspond to different gauge choices for $\epsilon^{\mu}$.
The advantage of this technique is that the $k^{\mu}$'s may be chosen to
simplify the combinations $\eps_i \cdot \eps_j$ and $\eps_i \cdot k_j$
which appear in $K_{\rm red}$. For the amplitude $M(1^-,2^-,3^+,4^+)$ choosing
$(k_i)=(p_4,p_4,p_1,p_1)$ then
$$
\eqalign{
k_4\cdot\eps_1 =& 0 ,\;\;\;  k_3\cdot\eps_1 = -k_2\cdot\eps_1 
\cr
k_4\cdot\eps_2 =& 0 , \;\;\; k_3\cdot\eps_2 = -k_1\cdot\eps_2
\cr
k_1\cdot\eps_3 =& 0 , \;\;\; k_4\cdot\eps_3 = -k_2\cdot\eps_3
\cr
k_1\cdot\eps_4 =& 0 ,\;\;\;  k_3\cdot\eps_4 = -k_2\cdot\eps_4
\cr}
\equn
$$
and 
$$
\eqalign{
\eps_i \cdot \eps_j =0 , \;\;\; i,j \neq 2,3 
\cr
\eps_2 \cdot \eps_3 = - { 2 \over t } k_1\cdot\eps_2 k_2\cdot\eps_3
\cr}
\equn
$$
With this simplification the entire kinematic expression is given by a 
kinematic factor
$$
\left(
\eps_1\cdot k_2
\eps_2\cdot k_1
\eps_3\cdot k_2
\eps_4\cdot k_2
\right)^2
\sim
 [stu M^{tree} ]
\equn
$$
times  a combination of Green's Functions. This simplifies the calculations
considerably.

\noindent
{\bf Step 1} The $M^{N=8}$ Amplitude

In the language of \cite{DunNor} we want the contribution of the form
$$
4[ C_V - 4 C_F ;  C_V - 4 C_F ]   
\equn
$$
for the Green's Functions.
Each of the combinations $C_V$ and $C_F$ are quadratic in the Feynman 
parameter polynomials however for the combination $C_V-4C_F$ two powers cancel
leaving just a constant.
For all three boxes this coefficient is simple as
$C_V - 4 C_F \rightarrow 1/2$ and we 
find the amplitude is just a sum over scalar box integrals with the overall
factor $[stu M^{tree} ]$.

\noindent
{\bf Step 2} The $M^{N=6}$.

In the language of \cite{DunNor} we want the contribution of the form
$$
-4[ C_V - 4 C_F ; C_F ] 
\equn$$

Applying the substitution rules 
we obtain, 
$$
[ stu M^{tree}/2 ] \times \left( I_{1234} [ f_1( a_i) ]
+I_{1243}[ f_2( a_i) ] +I_{1324}[f_3( a_i)]
\right) 
\equn
$$
where
$$
\eqalign{
 I_{1234} [ f_1( a_i) ] &= I_{1234}[ a_3 -a_3^2 +a_1a_3]
\cr
I_{1243}[ f_2( a_i) ] &= I_{1243}[ a_1 -a_1^2 +a_1a_3 -a_1a_2+a_2a_3]
\cr
I_{1324}[f_3( a_i)] & =
I_{1324}[ a_1 -a_1^2 -a_1a_3]
\cr}
\equn\label{EQpoly6}
$$
Note that this is true in any dimension. 

Reducing these integrals to scalar integrals by one's favorite
technique (our is that of ref.\cite{OneLoopInt}) we have
$$
\eqalign{
I_{1234}^{D}[ -a_1 +a_1^2 -a_1a_3 ]
={2 \over s} \times   (2-D/2)  I_{1234}^{D+2} 
+{2 \over st } I^{D}_2(t)  
\cr
I_{1234}^{D}[ 2a_1a_3 ]
={2s+2t(2-D/2)  \over su }I_{1234}^{D+2} 
-{2 \over su } I^{D}_2(s)
+{2 \over su } I^{D}_2(t)
\cr}
\equn
$$
which yields the amplitude
$$ 
-{1 \over s}I_{1324}^{D+2}
+ {D-4 \over 2 s}  \left(  I^{D+2}_{1234} + I^{D+2}_{1243}
- I^{D+2}_{1324} \right) 
\equn
$$
as has been observed before,~\cite{DimShift}, the amplitudes with less supersymmetry
have the dimensions of the box integrals shifted relative to the maximal supersymmetric
case.

\noindent
{\bf Step 3.} The $M^{N=4}$

In the language of \cite{DunNor} we want the contribution of the form
$$
2[ C_V - 4 C_F ; S ] 
\equn
$$
For the boxes we get 
$$
\eqalign{
I_{1234}&[
-a_3^2(1-a_3)^2
-a_2a_3a_4(1-a_3)
+2a_3^2(a_1+a_2)(a_1+a_4)
+a_3(a_2+a_4)/4]
\cr
I_{1243}&[(a_1+a_2)(a_3+a_4)(a_3a_1-a_1a_4-a_3a_2)
+(a_1a_4+a_3a_2)/4 ]\cr
I_{1324}&[
a_3^2(-a_4(a_1+a_4)-a_2(a_1+a_2)+a_2a_4)
-a_3a_2a_4 
+a_3(a_2+a_4)/4]
\cr}
\equn\label{EQpoly4}
$$
Reducing these integrals to scalar integrals we have
$$
{ (4-D)(2-D)  \over 4s^2} I^{D+4}(s,t) 
+{ ( 4-D)(2-D) \over s^2} I^{D+4}(s,u)
+{(D+2)D  \over 4s^2 }I^{D+4}(u,t)
-{1 \over 2 s^2} \left( I_2^D(t) +I_2^D(u) \right) 
\equn
$$

\noindent
{\bf Step 4}  The $M^{N=4*}$

Adding together the contributions to have
$$
M^{N=4*}
=M^{N=8}-4M^{N=6} +8M^{N=4}
\equn
$$
the combination is then just
$$
4[C_V - 4 C_F ; 4S+C_V  ]
\equn
$$
as one might deduce directly.

At this point we should discuss which string theory is relevant.  In
the original work on String-based rules the low energy limit of a
heterotic string theory was taken to obtain the rules. Later it was
realised that the same information could be encoded within 
a formalism where the fermionic sectors were dropped \cite{BernCompact}. 
One of the challenges in developing this technique was to break the link
between the string theory and the field theory limit allowing calculations
in more general field theories such as 
non-supersymmetric QCD to be performed~\cite{FiveGluon}. 

In our work we are focusing on $N=4$ supergravity which is the low
energy limit of heterotic string theory and as such the string based
rules ``recombine'' and simplify.  $N=4$ supergravity is of course the
field theory limit of two of the fundamental string theories namely
the type~I theory of open and closed strings and the Heterotic string
theory consisting of only closed strings. Both these string theories
contain gauge groups in the low energy limit which we decouple. In the
type~I theory the decoupling is extremely simple: the gauge particles
are in the open string sector so the contribution to a four graviton
scattering amplitude from the $N=4^*$ multiplet will come from the
torus and Klein bottle amplitudes only. For a heterotic theory one
must drop the contributions arising from the gauge particles arising from the
bosonic string.

\section{Unitarity Checks}
In this section we show how the $4-$graviton amplitude in $D=4$ may
be checked by combining unitarity and helicity
techniques~\cite{Cutting,SusyFour,Massive}.
This approach provides an
alternative method to obtain part of our results and at same time  a 
strong check of their consistency.

\begin{figure}[ht]
\centerline{\epsfxsize 2.2 truein \epsfbox{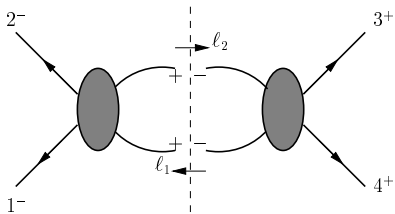}}
\caption[]{
\label{fCutE}
\small 
Helicity configuration for the $M(1^-,2^-,3^+,4^+)$ $s$-channel cut}
\end{figure}

\noindent
Let us begin with the $s$-channel cut of the four-point amplitude
$M(1^-,2^-,3^+,4^+)$ represented pictorially in fig. \ref{fCutE}. 
According to the Cutkosky rules, it is given by 
$$
 \left. {\rm Disc}~M^{1-loop}(1^-,2^-,3^+,4^+)
\right|_{s-cut}\!\!\!\!\!\!=
i \int\dlips\sum_{ {\rm internal}\atop {\rm particles}}
\,\Mtree(-\ell_1^+,1^-,2^-,\ell_2^+)\,\Mtree(-\ell_2^-, 3^+,4^+,\ell_1^- )
\equn
\label{FourPtCut}
$$ 
Here $dLIPS$ denotes the usual invariant Lorentz phase space and it
contains an additional symmetry factor $1/2$ to keep into account  
that 2 identical particles  are going through the cut. Since 
we wish to construct the entire amplitude, we observe that we can
replace  the phase space integral by the cut  of an unrestricted 
loop momentum integral
$$
\eqalign{
\left. M^{1-loop}(1^-,2^-,3^+,4^+)
\right|_{s-cut}=& \cr
&\!\!\!\!\!\!\!\!\!\!\!\!\!\!\!\!\!\!\!\!\!\!\!\!\!\!\!\!\!\!\!\!\!\!\!
\frac{1}{2}\int \frac{d^4\ell_1}{(2\pi)^4} 
\left.\!\!\!\sum_{ {\rm internal}\atop {\rm
particles}}\!\!\frac{i}{\ell_1^2}
\,\Mtree(-\ell_1^+,1^-,2^-,\ell_2^+)\,\frac{i}{\ell_2^2}
\Mtree(-\ell_2^-, 3^+,4^+,\ell_1^- )\right |_{s-cut\atop \ell_1^2=\ell_2^2=0}
}\equn\label{FourPtCut1}
$$
While eq. (\ref{FourPtCut}) includes only imaginary part, eq.
(\ref{FourPtCut1}) contains both real and imaginary parts. As
indicated, eq. (\ref{FourPtCut1}) holds only for those terms
with $s-$channel branch cut; terms without an $s-$channel cut 
require a separate determination. A very useful feature of this identity
is that we are free to use the on-shell conditions
$\ell_1^2=\ell_2^2=0$ to simplify the integrand: in fact only terms 
containing no cut in this channel would change.

\noindent
The only internal particles, which give a non-vanishing result in the
sum in eq. (\ref{FourPtCut1}), are gravitons.  Fermion contributions 
vanish because their helicity is not flipped by the graviton vertex 
implying that $M(g,g,\psi^+,\psi^+)=0$.
The same holds for vector and scalar amplitudes at tree level (taking
particle/antiparticle instead of positive/negative helicity in the
latter). ( On a more formal ground, this follows from SUSY and chiral
Ward identities~\cite{SWI}.)  Thus in our decomposition only the $N=8$
supermultiplet  will produce a non-vanishing $s-$cut, while the $N=6$ 
and $N=4$ supermultiplets will have only $t/u-$cuts. Let us focus on
the former case. 
%
\hskip 2truecm

\begin{figure}[ht]
\centerline{\epsfxsize 4 truein \epsfbox{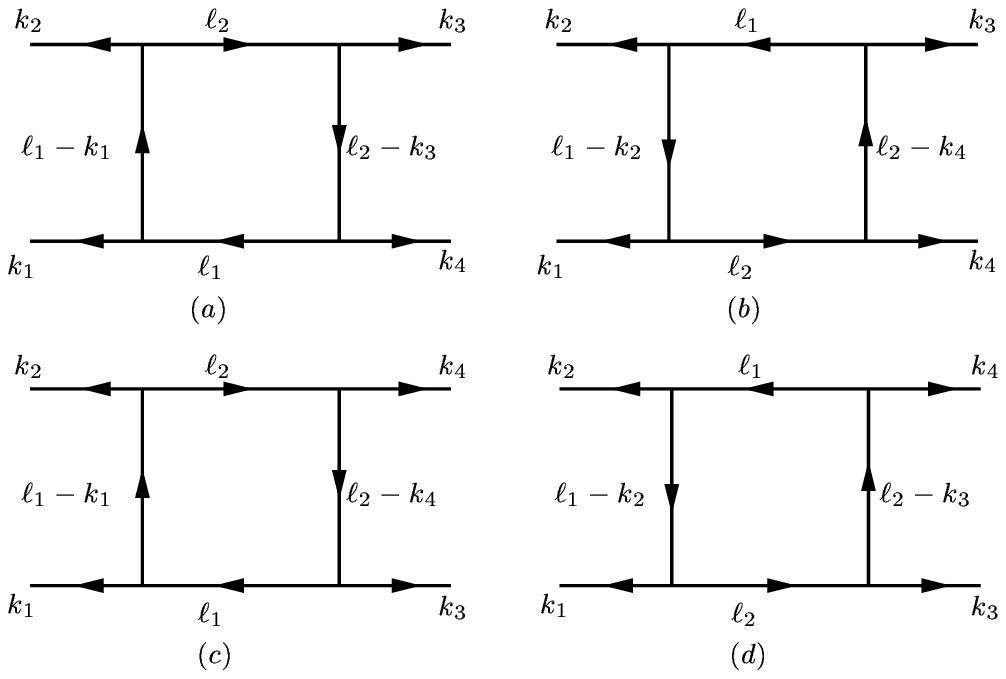}}
\vskip -.2 cm
\caption[]{
\label{Cuts8a}
\small 
Boxes which contribute to the $M^{1-loop}(1^-,2^-,3^+,4^+)$ calculation.}
\centerline {\small Note that (a) is
related to (b) and (c) to (d) by redefinition of loop momenta.}
\end{figure}

\noindent
In the standard helicity formalism (see {\it e.g.} \cite{Review}), the
tree amplitude has the form
$$
\Mtree(1^-,2^-,3^+,4^+)=i 
{\kappa^2 \over 4}
\biggl( { \spa1.2^4 \over \spa1.2\spa2.3\spa3.4\spa4.1
 } \biggr)^2
\times   { s t \over u }=
-i\frac{\kappa^2}{4}\left [\frac{1}{t}+\frac{1}{u}\right]\times 
\left[ s\frac{\spa1.2 [3 4]}{[1 2] \spa3.4}\right]^2.
\equn
\label{am1}
$$
The equivalence between the two forms of the amplitude can be easily
shown by exploiting the identities:$~[a b] \langle b a\rangle=2
(a,b)$ and 
$[a_1 a_2]
\langle a_2 a_3\rangle\cdots[a_{n-1} a_n] \langle a_n~a_1\rangle=1/2tr[(1+ \gamma_5) a_1\cdots a_n]\equiv
tr_+(a_1\cdots a_n).$ 
Inserting the r.h.s. of  eq. (\ref{am1}) into the sum appearing in
eq. (\ref{FourPtCut1}), we obtain
$$
\eqalign{-
{\kappa^4\over 16}
\left[ \frac{\spa1.2 [\ell_1 \ell_2]}{[1 2]\langle \ell_1
\ell_2\rangle}s\right]^2\left[\frac{1}{(\ell_1-k_1)^2}+\frac{1}{(\ell_1-k_2)^2}\right]\times 
\left[ \frac{[3 4]\langle \ell_1 \ell_2\rangle}{\spa3.4 [\ell_1
\ell_2]}s\right]^2\left[\frac{1}{(\ell_2-k_3)^2}+\frac{1}{(\ell_2-k_4)^2}\right]
}
$$
which we can rearrange to be,
$$
i {\kappa^2\over 4} s t u
M^{tree}(1^-,2^-,3^+,4^+)\times   \left[\frac{1}{(\ell_1-k_1)^2}+\frac{1}{(\ell_1-k_2)^2}\right] \left[\frac{1}{(\ell_2-k_3)^2}+\frac{1}{(\ell_2-k_4)^2}\right]
\equn\label{CUTaa}
$$
where we have factorized out the kinematical factor given by the tree
amplitude. Substituting this result into eq. (\ref{FourPtCut1}) and 
expanding the products, we can interpret each term as coming from a 
box integral  depicted in Figure (\ref{Cuts8a}). 
%
For example the product of the first two propagator leads to the integral
$$
\int {d^4l_1 \over (2 \pi)^{4}}
{1\over \ell_1^2(\ell_1-k_1)^2\ell_2^2(\ell_2-k_3 )^2},\equn
$$
which corresponds to the box integral $(a)$ in fig. (\ref{Cuts8a}). 
$$
\left. M^{1-loop}(1^-,2^-,3^+,4^+)\right |_{s-cut} =
\left.  \left ({\kappa\over 2}\right)^2 [ s t
u\Mtree(1^-,2^-,3^+,4^+)]
(4\pi)^{-2}\left[I^4_4 (s,t)+I^4_4(s,u)\right]\right |_{s-cut},
\equn
\label{scut}
$$
which is manifestly in  agreement with the general result given in eq.
(2.6) for the $N=8$ contribution to the amplitude. [ $I_4^4(t,u)$ is
absent in (\ref{scut}) because it does not contain any $s-$cut. Its
presence can be detected by looking at the $t/u-$cuts.] 

\begin{figure}[ht]
\centerline{\epsfxsize 2.2 truein \epsfbox{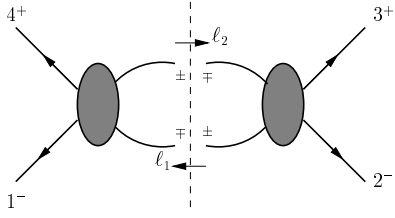}
}
\vskip -.2 cm
\caption[]{
\label{fCutG}
\small 
Helicity configurations for the $M(1^-,2^-,3^+,4^+)$ $t$-channel cut}
\end{figure}

The $t$- and $u$-channel cuts require essentially the same calculations
since the helicity configuration are the same in both cases. We shall focus
on the $t$-channel case and deduce the $u$-channel result from this by
permuting $(1\leftrightarrow 2)$. In this case the possible helicity
configurations of the trees are those represented in fig.~\ref{fCutG}.
Since helicity is no longer flipped on either tree, the cut receives a
non-vanishing contribution from each of the supermultiplet. 
%

\noindent
The contribution to the $t$-cut from states in the self-conjugate 
supermultiplet with  $N$ supersymmetries  is given by 
$$
\eqalign{
{\rm Disc} M^{1-loop}&\left.(1^-,2^-,3^+,4^+)\right |_{t-cut}=\cr
& i \int~\dlips
\sum_{J\in N\atop \rm supermultiplet}
 \Mtree(4^+,1^-,\ell_2^{J-},-\ell_1^{J+})\ 
 \Mtree( \ell_1^{J-},-\ell_2^{J+}, 2^-,3^+ )}
\equn\label{tcut}
$$
The above sum can be performed by recalling that all amplitudes  are
related, via supersymmetric Ward identities (see refs.~\cite{SWI}), to
the scalar one according to the relations
$$
\eqalign{
\Mtree(g^-,g^-,g^+,g^+) & = y^4 \Mtree(g^-,\phi^-,\phi^+,g^+)
\cr
\Mtree(g^-,\psi^-,\psi^+,g^+) & = y^3 \Mtree(g^-,\phi^-,\phi^+,g^+)
\cr
\Mtree(g^-,A^-,A^+,g^+) & = y^2 \Mtree(g^-,\phi^-,\phi^+,g^+) 
\cr
\Mtree(g^-,\Lambda^-,\Lambda^+,g^+)  &= y\Mtree(g^-,\phi^-,\phi^+,g^+)
\cr}
\equn\label{rel18}
$$
where  $y=\spa1.2/\spa1.3$ and $g,~\psi,~A,~\Lambda$ and $\phi$ denote
the graviton, the gravitino, the vector, the spin $1/2$ and the scalar
respectively. Using, in fact, the relations
(\ref{rel18}) we can rearrange eq. (\ref{tcut}) to be, 
$$\eqalign{ 
{\rm Disc}M^{1-loop}(1^-,2^-,&\left.3^+,4^+)\right |_{t-cut}=\cr
&i \int \dlips \
 \Mtree(4^+,1^-,\ell_2^{S-},-\ell_1^{S+})\,
\, \Mtree( \ell_1^{S-},-\ell_2^{S+}, 2^-,3^+ )~\rho_N}
\equn
$$
where $ \rho_N = (x - x^{-1})^N $ with
$$
x^2 = { \spa1.{\ell_2}\spa2.{\ell_1} \over \spa1.{\ell_1}\spa2.{\ell_2} }.
\equn\label{tcut1}
$$
The term $\rho_N$ is a kinematical factor that encodes the sum over the
different particles, while $\Mtree$ is, in this case, the amplitude for two
scalars into two gravitons, whose  explicit form can be obtained
by combining eq.  (\ref{rel18}) with eq. (\ref{am1}).

\noindent
If we restrict ourselves to choices with even $N=2m$, we can simplify
$ \rho_{2m}$ as follows
$$
\eqalign{
\rho_{2m} &= (x-x^{-1})^{2m} = { (x^2-1)^{2m} \over x^{2m} }
={ (\spa1.{\ell_2} \spa2.{\ell_1}-\spa1.{\ell_1}\spa2.{\ell_2} )^{2m}
\over (\spa1.{\ell_2}\spa2.{\ell_1}\spa1.{\ell_1}\spa2.{\ell_2})^m }=
\cr
& ={\spa1.2^{2m} \spa{\ell_1}.{\ell_2}^{2m}
\over (\spa1.{\ell_2}\spa2.{\ell_1}\spa1.{\ell_1}\spa2.{\ell_2})^m }
=\rho_8 \left ({ \spa1.{\ell_2}\spa2.{\ell_1}\spa1.{\ell_1}\spa2.{\ell_2}
\over \spa1.2^2 \spa{\ell_1}.{\ell_2}^{2}}\right)^{4-m}=\cr
&= \left({\,  \tr_+(4 \ell_1 1 2 \ell_1 3) \over  t\, s^2}\right)^{4-m}\rho_8.\cr}
\equn
$$
where we have used the well-known identity $\langle a  b\rangle 
\langle c d\rangle  + \langle a  c\rangle \langle d  b\rangle +
\langle a  d\rangle \langle b c\rangle =0$.

\noindent
It remains to compute the combination
$\rho_8 ~\Mtree(4^+,1^-,\ell_2^{S-}, -\ell_1^{S+})\, \Mtree(
\ell_1^{S-},-\ell_2^{S+}, 2^-,3^+ )$ which corresponds to 
the contribution of the $N=8$ self-conjugate supermultiplet 
to the $t-$cut. Inserting the explicit expression for the 
amplitudes and $\rho_8 $ we find
$$
-\left(  { \kappa^2\over 4}\right)^2t^2
{ \spa1.2^8 \over \spa2.3^2\spa4.1^2 }
{ (k_1\cdot   \ell_2) \over (k_1\cdot   \ell_1) }
{ (k_2\cdot   \ell_2) \over (k_2\cdot   \ell_1) }
{
\spa{\ell_1}.{\ell_2}^4 \over
             ( \spa1.{\ell_2}\spa{\ell_1}.4\spa{\ell_2}.2\spa3.{\ell_1} )^2 },
\equn
$$
which in turn can be rewritten as
$$\eqalign{
i \left( {\kappa^2\over 4}\right)
[ s t uM^{tree}(4^+,1^-,2^-,3^+)] 
{ t^2 \over 16 (k_1\cdot   \ell_1)(k_2\cdot   \ell_1)
                  (k_4\cdot   \ell_1 )(k_3\cdot   \ell_1 ) }}
\equn
$$
Here, as in the case of the $s-$cut, we have factorized out the tree amplitude
and reduced everything to scalar products.

\noindent
Therefore, we are led to write the following master-formula that
encompasses all the possible contributions to the $t-$channel coming
from a self-conjugate supermultiplet in $D=4$

\noindent

$$\eqalign{
{\rm Disc} M^{1-loop}(1^-,2^-&,3^+,4^+)\left. \right |_{t-cut}=
i \left( { \kappa^2\over 4}\right)
[ s t u M^{tree}(1^-,2^-,3^+,4^+)]\times   \cr
& i \int\dlips\,{ t^2 \over 16 (k_1\cdot   \ell_1)(k_2\cdot   \ell_1)
                  (k_4\cdot   \ell_1 )(k_3\cdot   \ell_1 )}\,
\left({  \tr_+(4 \ell_1 1 2 \ell_1 3)
\over  t\, s^2}\right)^{4-m} }.
\equn
\label{gen1}
$$
The remarks made after eq. (\ref{FourPtCut1}) are also valid here. In 
particular we are free to turn the scalar products appearing in the  
denominator of eq. (\ref{gen1}) into propagators, when it is convenient.

\noindent
The $N=8$ contribution is selected by choosing $m=4$. In this case, using the same
procedure adopted for the $s-$channel, we can turn eq. (\ref{gen1}) into a
sum of boxes. We find a very similar result except that $s\leftrightarrow t$.

\begin{figure}[ht]
\centerline{\epsfxsize 6 truein \epsfysize 7 truecm\epsfbox{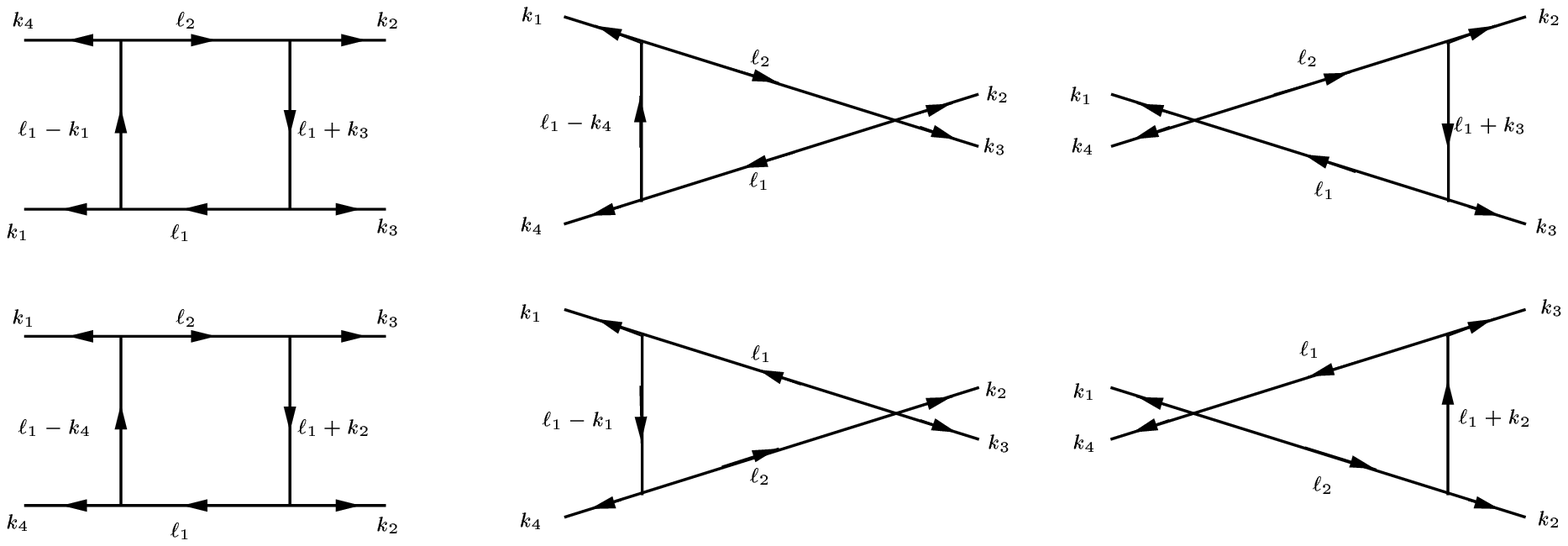}
}
\vskip -.2 cm
\caption[]{
\label{fCutH}
\small 
Boxes and triangles contributing to $M(1^-,2^-,3^+,4^+)$ $t$-channel cut}
\end{figure}

\noindent
By posing $m=3$, we are led to compute the $N=6$ contribution. The first
step is to expand the trace in the integrand and to simplify the common
factors with the denominator
$$
\eqalign{
&\frac{\tr_+(4 \ell_1 1 2 \ell_1 3)}{16(k_1\cdot   \ell_1)(k_2\cdot   \ell_1)
                  (k_4\cdot   \ell_1 )(k_3\cdot   \ell_1 )  }
=\frac{\tr(4 \ell_1 1 2 \ell_1 3)}{32(k_1\cdot   \ell_1)(k_2\cdot   \ell_1)
                  (k_4\cdot   \ell_1 )(k_3\cdot   \ell_1 ) }
=\cr
&=
-\frac{1}{8}\left[ t\left(\frac{1}{(\ell_1\cdot   k_1)(\ell_1\cdot   k_4)}+\frac{1}
{(\ell_1\cdot   k_2)(\ell_1\cdot   k_3)}\right)
-u\left(\frac{1}{(\ell_1\cdot   k_1)(\ell_1\cdot   k_3)}+\frac{1}{
(\ell_1\cdot   k_2)(\ell_1\cdot   k_4)}\right)\right]=\cr
&=-\frac{1}{8}  \left[ 2\left(\frac{1}{(\ell_1\cdot   k_1)}+
        \frac{1}{(\ell_1\cdot   k_4)}-\frac{1}{(\ell_1\cdot   k_2)}
-\frac{1}{(\ell_1\cdot   k_3)}\right)
-u\left(\frac{1}{(\ell_1\cdot   k_1)(\ell_1\cdot   k_3)}+\frac{1}{
(\ell_1\cdot   k_2)(\ell_1\cdot   k_4)}\right)\right]\cr}
\equn
\label{cio}
$$
By rewriting the denominators in the last line of eq. (\ref{cio})
as propagators,
$$
\frac{1}{2}\left[ \frac{1}{(\ell_1-k_1)^2}
+\frac{1}{(\ell_1-k_4)^2}+\frac{1}{(\ell_1+k_2)^2}+\frac{1}{(\ell_1+k_3)^2}
-\frac{u}{(\ell_1-k_1)^2(\ell_1+k_3)^2}-
\frac{u}{(\ell_1-k_4)^2(\ell_1+k_2)^2}\right]\equn
$$
we can finally turn the  phase space integral in eq. (\ref{gen1}) 
into the $t-$cut of an unrestricted loop integral. As in the case of the 
$s-$channel, this is achieved through the substitution 
$$
i\int\dlips \to \left. \int \frac{d\ell_1}{(2\pi)^4}\frac{i}{\ell_1^2}
\frac{i}{\ell_2^2} \ \ \ \right|_{t-cut}.\equn
$$
The final result is a combination of four triangles $I^4_3$ 
and two boxes  $I^4_4$ and it is represented pictorially 
in fig. \ref{fCutH}.
In terms of the integral functions, we have

$$
\eqalign{
 M^{1-loop}&(1^-,2^-,3^+,4^+) \bigg|_{t-cut}=
\cr
& 
(-\frac{1}{2})\left( \frac{\kappa}{2} \right)^2 [ s t
u\Mtree(1^-,2^-,3^+,4^+)](4\pi)^{-2} \frac{u t}{s^2}
\left( \frac{2}{u} I^4_3(t)+\frac{2}{t} I^4_3(u)
+I^4_4(t,u) \right)
\bigg|_{t-cut}
\cr}
\equn
$$

where we have already added the term $\displaystyle{\frac{2}{t}
I^4_3(u)}$, whose presence can be detected by looking at the $u-$channel.
Then using the identity 
$$
I_4^6(t,u)=-\frac{t u}{2s} \left ( I^4_4(t,u)+\frac{2}{t}I^4_3(u)+
\frac{2}{u}I^4_3(t)\right)
\equn
$$
we can recast the above combination of boxes and triangles integrals 
as a  $6-$dimensional box
$$
\left. M^{1-loop}(1^-,2^-,3^+,4^+)\right |_{t-cut}=\left.
\left(\frac{\kappa}{2}\right)^2 
[ s t u\Mtree(1^-,2^-,3^+,4^+)](4\pi)^{-2} \frac{1}{s} I^6_4(t,u)
\right |_{t-cut}.
\equn
$$
This is exactly the result expected from formula (2.6).

The $N=4$ contribution is more lengthy, but it can be computed along the
same lines. We find again agreement with eq. (2.6). However in this last
case some caution is in order: while the $N=8$ and $N=6$ amplitudes were
cut-constructible, namely their exact one-loop expression could be deduced
using unitarity~\cite{Cutting,SusyFour,Massive}, the $N=4$ is not and
only the cut expressions are guaranteed to coincide.

\vfill\eject

\section{ Form of the Counterterms and Factorisation of the Divergences}

The aim of the present appendix is to express each factor appearing in
the $1/\epsilon$ poles of the $D=8$ and $D=10$ amplitudes 
 in terms of Lorentz
 invariants  in the product of four gauge field--strengths $F_{ab}$ and their
derivatives. One of the advantages of this analysis was to suggest 
a natural (and correct) form  for the
$\partial^2 R^4$ invariants entering the expression of our
counterterm in $D=10$. This would have been otherwise a hard task
since the number of independent $\partial^2 R^4$ invariants in $D=10$
dimensions which do not vanish on--shell is more than 30
\cite{Fulling}.
 At the end we shall briefly comment on a possible
superstring interpretation of the observed factorization of the
counterterms,
which actually extends to the complete 1-loop amplitude as well.\par
It is useful to express the Riemann tensor in terms of the symmetrized
product of two field--strengths:
\begin{eqnarray}
R_{ab,cd}&=& \frac{1}{2}\left(F_{ab} \bar{F}_{cd}+\bar{F}_{ab}
F_{cd}\right)\nonumber\\
F_{ab} &\equiv &
k_{\left[a\right.}\epsilon_{\left. b\right]}\,;\,\bar{F}_{ab} \equiv
k_{\left[a\right.}\bar{\epsilon}_{\left. b\right]}\nonumber
\end{eqnarray}
the vector polarizations $\epsilon_a$ and $\bar{\epsilon}_b$ being
related to the graviton polarization $\epsilon_{ab}$ through a
traceless symmetrized product: $\epsilon_{ab}=\epsilon_{(a}\bar{\epsilon}_{b)}$.
As shown in Sections 3 and 4, the $D=8$ counterterms we find are of
the form $K_1 \times K_j$ and similarly  the $D=10$ counterterms have
the factorized form $K_1 \times L_j$ ($j=1,2$). As we shall show
below, each of the factors can be expressed as a suitable invariant in
the product of four vector field--strengths $F_{ab}$ and their
derivatives. In particular, in the $D=8$ case 
we find that the  factors  $K_i$ are two independent invariants
constructed contracting
suitable combinations of the tensors
$t_{(12)}$ and $t_{(48)}$ with four gauge field strengths
$F_{ab}$. \\
If we define:
$$
\eqalign{
N_1 =& -\frac{1}{12}\, t_{(12)}\cdot F^4=( F_{ab} F^{ab} ) ( F_{cd}F^{cd} )
\cr
N_2 =& \frac{1}{48}\, t_{(48)}\cdot F^4=F_{a}{}^b F_b{}^c F_c{}^d F_d{}^a
\cr}
\equn
$$
 we can express $K_1$ and $K_2$ as follows:
$$
\eqalign{
K_1 =&{ 1 \over 16} \Bigl( N_1 -4N_2 ) =-{1\over 96} t_8\cdot F^4  
\cr
K_2 =&-{ 1 \over 16} \Bigl( 5N_1 +4N_2 \Bigl) 
\cr }
\equn
$$
In $D=10$ we find that the factors $L_i$  in the expression of the
counterterm are
related to invariants involving $\partial^2 F^4$ terms. 
There are only three independent such terms which do not vanish on--shell 
\cite{tsetse} which have the following form:
$$
\eqalign{
J_1 =& (\partial_m F_{ab} \partial^m F^{ab} ) ( F_{cd}F^{cd} )
\cr
J_2 =&\partial_m F_{a}{}^b \partial^m F_b{}^c F_c{}^d F_d{}^a
\cr
J_3 =& 
\partial_a F_{m}{}^n \partial_b F^m{}_n  F_a{}^p F_b{}^p
\cr}
\equn
$$
In terms of the above invariants, $L_1$ and $L_2$ are expressed as follows: 
$$
\eqalign{
L_1 =& \Bigl( 5J_1 +2J_2 -2J_3 \Bigr)\cr
L_2 =&\Bigl( -{J_1 \over 2} +2J_2 \Big)\cr}
\equn
$$
It is useful to express the product $K_1\times J_k$ in the basis of
the invariants $S_I$ ($I=1,\dots, 10$):
$$
\eqalign{
K_1\times J_1 &\longrightarrow  
{ -{ 1 \over 8} } \Bigl( 
-4S_{1}+32S_{2}-8S_{4}+16S_{5} \Bigr)
\cr
K_1\times J_2  &\longrightarrow  
{ -{ 1 \over 8} } \Bigl( 
-4S_{2}-4S_{3}+4S_{5} -8S_{6} +16S_{7} +8S_{8} \Bigr)
\cr
K_1\times J_3  &\longrightarrow 
{ -{ 1 \over 8} } \Bigl( 
-S_{1}+40S_{3}+10 S_{4} -12S_{5} +80S_{6} +32S_{7} -80S_{8} 
+24S_9 -95 S_{10} \Bigr)
\cr}
\equn
$$
So that, multiplying by $K_1$ and using eqs. (A.4) we find the
expression given in section 4 on the $D=10$ counterterms ($N=6$ and
$N=4$ respectively):
$$
\eqalign{
K_1 L_2 &\longrightarrow
- { 1 \over 4}S_{1}+3 S_{2}+{S_{3} } -{S_{4} \over 2} 
+0.S_{5} +2S_{6} -4 S_{7} -2S_{8}
\cr
K_1 L_1  &\longrightarrow  
-{1 \over 8} \Bigl(
-18S_{1} +152S_{2} -88{S_{3} } -60S_{4}
+112S_{5} -176S_{6} -32 S_{7} +176S_{8}-48S_{9}+190S_{10}
\Bigr)
\cr}
\equn
$$

We found it useful also to express $K_1\times J_3$ in  a different
basis of invariants $M_l$:
$$
K_1\times J_3  \longrightarrow 
4M_1 -16M_2 -16M_3 +4M_4 -16 M_5 +4M_6 
\equn
$$
Where 

\def\Mtensor#1#2#3#4{\partial_{a} R_{ef{#1}{#2}} \partial_{d} 
R_{ef{#3}{#4}}R_{dh{#3}{#4}} 
R_{ha{#1}{#2}}}

$$
\eqalign{
\cr  
M_1 &=
\partial_{a} R_{eflm} \partial_{d} R_{efno}R_{dhno} R_{halm}
\cr
M_2 &=
\partial_{a} R_{eflm} \partial_{d} R_{efno}R_{dhmo} R_{haln}
\cr
M_3 &=
\partial_{a} R_{eflo} \partial_{d} R_{efno}R_{dhnm} R_{halm}
\cr
M_4 &=
\partial_{a} R_{eflm} \partial_{d} R_{efno}R_{dhlm} R_{hano}
\cr
M_5 &=
\partial_{a} R_{eflo} \partial_{d} R_{efno}R_{dhlm} R_{hanm}
\cr
M_6 &=
\partial_{a} R_{efmo} \partial_{d} R_{efmo}R_{dhln} R_{haln}
\cr}
\equn
$$

The expression of $K_1\times J_3$ in terms of the tensors $M_l$ is
related through (torsion) Bianchi identities to its expression in
terms of the $S_i$. The first two equations in (A.5) expressing $K_1\times J_{1,2}$
in terms of $S_I$ ($I=1,\dots, 8$) and (A.7) expressing
$K_1\times J_3$ in terms of the $M_l$ ($l=1,\dots, 6$) can be derived 
directly by writing the three $K_1\times J_i$ in the following more compact
way:
\begin{eqnarray}
K_1\times J_i &=& d_{J_i}t_8\cdot \partial^2 R^4\nonumber\\
d_{J_i}t_8\cdot \partial^2 R^4
&\equiv &
 d_{J_i}^{a_1a_2a_3a_4a_5a_6a_7a_8a_9a_{10}}
t_8{}_{b_1b_2b_3b_4b_5b_6b_7b_8} 
\partial_{a_1} R_{a_2a_3}{}^{b_1b_2}
 \partial_{a_4} R_{a_5a_6}{}^{b_3b_4}
R_{a_7a_8}{}^{b_5b_6}
R_{a_9a_{10}}{}^{b_7b_8}\nonumber
\end{eqnarray}
and the $10$--tensor 
$d_{J_i}$ is defined as follows:
$$
\eqalign{
d_{J_1}^{a_1a_2a_3a_4a_5a_6a_7a_8a_9a_{10}}
&= \delta^{ a_1a_4}\delta^{a_2a_5}
\delta^{a_3a_6}\delta^{a_7a_9}\delta^{a_8a_{10}}
\cr
d_{J_2}^{a_1a_2a_3a_4a_5a_6a_7a_8a_9a_{10}}
&= \delta^{a_1a_4}\delta^{a_2a_{10}}\delta^{a_3a_5}
\delta^{a_6a_7}\delta^{a_8a_9}
\cr
d_{J_3}^{a_1a_2a_3a_4a_5a_6a_7a_8a_9a_{10}}
&=-\delta^{a_1a_{10}}\delta^{a_2a_5}\delta^{a_3a_6}
\delta^{a_4a_7}\delta^{a_8a_9}
\cr}
\equn
$$
The observed factorization of the $N=1$ and $N=2$ counterterms is consistent
with the interpretation of these theories as low--energy limits of
suitable closed superstring theories (heterotic and Type II respectively).
Indeed, as it was shown above, each factor in the expression of the
counterterms can be written in terms of invariants within suitable gauge
theories.
These gauge theories  may be
thought of as describing the low--energy excitations of the open string
theories associated with the {\it left} (L) and {\it right} (R) movers of some
closed string. From this point of view the factorization of the
counterterms (and of the whole amplitudes) would seem quite natural: 
it is well known that the free Fock space of a closed string theory  is the
tensor product of 
the two open string Fock spaces  corresponding to the (L) and(R) movers,
moreover, retrieving the supergravity amplitudes from the  $\alpha^\prime$
expansion of closed string amplitudes (that is using
superstring theory as a regulator of the effective supergravity
theory), up to one--loop order the contributions from the two sectors 
are expected to factorise. However we wish to emphasize that the philosophy
underlying our work is not to go from superstring theory "downwards", that
is to study supergravity as its effective low energy theory, but on the
contrary to move from field theory "upwards" and to study its effective
action within a QFT framework, using dimensional reduction as a 
regularisation scheme, instead of string theory. From the field
theory  point of view  the observed factorization is a highly non trivial
result, consistent with the interpretation of the  supergravities
 with $32$ and $16$ supercharges as low energy theories of type II and
heterotic superstring theories respectively.
In particular the different forms of the counterterms in  the maximal and
non maximal cases may be related, in this perspective, to different
 supersymmetry constraints holding on the two sectors of the closed string
theory. More specifically our results seem to suggest that in an open string
sector the only possible invariants
in four $F_{ab}$ consistent with an $N=1$ supersymmetry have to be
constructed  by saturating the indices of the field strengths with a $t_8$
tensor. This condition would restrict the possible invariants (non
vanishing on--shell) 
in four $F_{ab}$ and at most two derivatives, to just one possibility, namely
$K_1\propto t_8\cdot F^4$. Indeed if we interpret the maximal supergravity 
as the low energy limit of type II superstring, which has an $N=1$
supersymmetry on both the (L) and (R) sectors, according to the above
positions the only on--shell non vanishing counterterm is the one found in
$D=8$ (with no derivatives), which is indeed proportional to $K_1\times K_1$.
As far as the theory with $16$ supercharges is concerned, it may be
interpreted as the low energy effective theory of the heterotic
superstring, which has an $N=1$ supersymmetry on one sector (L) and an $N=0$
on the other (R). Again, consistently with our results and the above considerations, we can associate the $K_1$ factor of the corresponding counterterms
in $D=8$ and $D=10$ with the sector (L) constrained by supersymmetry.

\vfill\eject


\small
\begin{thebibliography}{99}


\bibitem{GSW}
M.B.\ Green, J.H.\ Schwarz and E. Witten,
{\it Superstring Theory} (Cambridge University Press, 1987).

\bibitem{CJS} 
E.\ Cremmer, B.\ Julia and J.\ Scherk, Phys.\ Lett.\ {\bf B76}:409 (1978). 

\bibitem{ExtendedSugra}
E.\ Cremmer and B.\ Julia, 
Phys.\ Lett.\ {\bf B80}:48 (1978);Nucl.\ Phys.\ {\bf B159}:141 (1979).   

\bibitem{BDDPR}
Z. Bern, L. Dixon, D.C. Dunbar, M.\ Perelstein and J. Rozowski, 
Nucl.\ Phys.\ {\bf B530}:401 (1998) [hep-th/9802162];
[hep-th/9809163].

\bibitem{DS}
S. Deser and D.\ Seminara, 
Phys.\ Rev.\ Lett.\ {\bf 82} (1999) 2435, [hep-th/9812136]. 

\bibitem{NeightCounter}
S.\ Deser, J.H.\ Kay and K.S.\ Stelle,
Phys.\ Rev.\ Lett.\ {\bf 38}:527 (1977)\\
R.E. Kallosh, Phys.\ Lett.\ {\bf B99}:122 (1981)\\
P.S. Howe, K.S. Stelle and P.K. Townsend, Nucl.\ Phys.\ {\bf B191}:445 (1981).

\bibitem{Heterotic}
D.J.~Gross, J.A.~Harvey, E.~Martinec and R.~Rohm,
Phys. Rev. Lett. {\bf 54} (1985) 502.

\bibitem{J85}
B.~Julia,
Proc. AMS-SIAM Summer Seminar on Applications of Group Theory in 
Physics and Mathematical Physics, Chicago 1982, eds. M. Flato, P. Sally \& 
G. Zuckerman, Lectures in Applied Mathematics, Vol. 21 (1985) p.335.


\bibitem{J9805}
B.~Julia,
in Proceedings of Carg\`ese NATO ASI: "Strings, branes
and dualities"  1997 ed. P. Windey et al.

[hep-th/9805083].
\bibitem{Helg}
S. Helgason, Differential Geometry, Lie groups and symmetric spaces 
Academic Press 1978.

\bibitem{Long}
Z. Bern and D.A.\ Kosower, Nucl.\ Phys.\ {\bf B379}:451 (1992).

\bibitem{Review}
Z. Bern, L. Dixon and D.A. Kosower, 
Ann.\ Rev.\ Nucl.\ Part.\ Sci.\ {\bf 46}:109 (1996) [hep-ph/9602280]. 

\bibitem{BDS}
Z. Bern, D.C. Dunbar and T. Shimada, 
Phys.\ Lett.\ {\bf B312}:277 (1993) [hep-th/9307001].


\bibitem{DunNor}
D.C. Dunbar and P.S. Norridge,  
Nucl.\ Phys.\ {\bf  B433}:181 (1995) [hep-th/9408014]; 
 Class.\ Quant.\ Grav.\ {\bf 14}:351 (1997)
[hep-th/9512084].

\bibitem{DRED}
W. Siegel, Phys.\ Lett.\ 84B:193 (1979);\\\
D.M.\ Capper, D.R.T.\ Jones and P. van Nieuwenhuizen, Nucl.\ Phys.\
B167:479 (1980).

\bibitem{GSB}
 M.B.\ Green, J.H.\ Schwarz and L.\ Brink, Nucl.\ Phys.\ {\bf B198}:474
 (1982).

\bibitem{SpinorHelicity}
P.\ De Causmaecker, R.\ Gastmans, W.\ Troost and T.T.\ Wu,
Phys.\ Lett.\ {\bf B105}:215 (1981), Nucl.\ Phys.\ {\bf B206}:53 (1982);\\
R.\ Kleiss and W.J.\ Stirling, Nucl.\ Phys.\ {\bf B262}:235 (1985);\\
J.F.\ Gunion and Z.\ Kunszt, Phys.\ Lett.\ {\bf B161}:333 (1985);\\
Z. Xu, D.-H.\ Zhang and L. Chang, Nucl.\ Phys.\ {\bf B291}:392 (1987); \\ 
M.L.~Mangano and S.J.~Parke,
Phys. Rept. {\bf 200} (1991) 301.

\bibitem{BGK}
F.A.~Berends, W.T.~Giele and H.~Kuijf,
Phys. Lett. {\bf B211} (1988) 91; \\
H.T.~Cho and K.L.~Ng,
Phys. Rev. {\bf D47} (1993) 1692; \\ 
D.~Spehler and S.F.~Novaes,
Phys. Rev. {\bf D44} (1991) 3990.

\bibitem{BernGrant}
Z.~Bern and A.K.~Grant,
Phys. Lett. {\bf B457} (1999) 23
[hep-th/9904026].

\bibitem{Tani}
N.\ Sakai and Y.\ Tanii,
Nucl.\ Phys.\ {\bf B287} (1987) 457.

\bibitem{Tsey}
A.\ Tseytlin, Nucl. Phys. {\bf B467} (1996) 383.


\bibitem{GreenVH}
M.B.~Green, M.~Gutperle and P.~Vanhove,
Phys. Lett. {\bf B409} (1997) 177
[hep-th/9706175].

\bibitem{RussoTsey}
J.G.\ Russo and  A.A.\ Tseytlin, 
Nucl.\ Phys.\ {\bf B508}:245 (1997) 
[hep-th/9707134]. 


\bibitem{DeightSugra}
M.~Awada and P.K.~Townsend,
Phys. Lett. {\bf B156} (1985) 51; \\
A.~Salam and E.~Sezgin,
Nucl. Phys. {\bf B258} (1985) 284.

\bibitem{BelRobinson}
I. Robinson, unpublished;\\
L.\ Bel, Acad. Sci. Paris, Comptes Rend. {\bf 247} :1094 (1958) ;
{ \bf 248} :1297 (1959).

\bibitem{dR}
M.~de Roo, H.~Suelmann and A.~Wiedemann,
Nucl. Phys. {\bf B405} (1993) 326
[hep-th/9210099].


\bibitem{Fulling}
S.A.\ Fulling, R.C.\ King, B.G.\ Wybourne and C.J.\ Cummins
Class.\ Quant.\ Grav.\ {\bf 9}:1151  (1992).

\bibitem{BernCompact}
Z.~Bern,
Phys.\ Lett.\ {\bf B296} (1992) 85.

\bibitem{FiveGluon}
Z.~Bern, L.~Dixon and D.A.~Kosower,
Phys.\ Rev.\ Lett.\ {\bf 70} (1993) 2677

\bibitem{tsetse}
O.D. Andreev, A.A. Tseytlin, Nucl. Phys. {\bf B311} (1988/89) 205.

\bibitem{WorldLine}
M.J.~Strassler,
Nucl.\ Phys.\ {\bf B385} (1992) 145
hep-ph/9205205 ; \\ 
M.G.~Schmidt and C.~Schubert,
Phys.\ Lett.\ {\bf B331} (1994) 69
hep-th/9403158;
Phys.\ Lett.\ {\bf B318} (1993) 438
hep-th/9309055.

\bibitem{Kikuchi1986}
Y.~Kikuchi, C.~Marzban and Y.J.~Ng,
Phys. Lett. {\bf B176} (1986) 57.

\bibitem{Cai:1987sa}
Y.~Cai and C.A.~Nunez,
Nucl. Phys. {\bf B287} (1987) 279.

\bibitem{Ellis:1988dc}
J.~Ellis, P.~Jetzer and L.~Mizrachi,
Nucl. Phys. {\bf B303} (1988) 1.

\bibitem{BernDunbar}
Z. Bern and  D.C. Dunbar, Nucl.\ Phys.\ {\bf B379}:562 (1992). 

\bibitem{OneLoopInt}
Z. Bern, L. Dixon and D.A. Kosower, Nucl.\ Phys.\ {\bf  B412}:751 (1994)
[hep-ph/9306240].

\bibitem{DimShift}
Z. Bern, L. Dixon, D.C. Dunbar and D.A. Kosower, 
Phys.\ Lett.\ {\bf B394}:105 (1997) [hep-th/9611127]. 

\bibitem{Cutting}
L.D.\ Landau, Nucl.\ Phys.\ {\bf 13}:181 (1959);\\
S. Mandelstam, Phys.\ Rev.\ {\bf 112}:1344 (1958), 115:1741 (1959);\\
R.E.\ Cutkosky, J.\ Math.\ Phys.\ {\bf 1} :429 (1960).

\bibitem{SusyFour}
Z. Bern, L. Dixon, D.C. Dunbar and D.A. Kosower,
Nucl.\ Phys.\ {\bf B425}:217 (1994) [hep-ph/9403226];
Nucl.\ Phys.\ {\bf B435}:59 (1995) [hep-ph/9409265];
[hep-ph/9706477]; 
Nucl.\ Phys.\ Proc.\ Suppl.\ {\bf 39BC} (1995) 146 [hep-ph/9409214].


\bibitem{Massive}
Z. Bern and A.G.\ Morgan, Nucl.\ Phys.\ {\bf B467} :479 (1996) [hep-ph/9511336].


\bibitem{SWI}
M.T.\ Grisaru, H.N.\ Pendleton and P.\ van Nieuwenhuizen,
Phys.\ Rev.\ {\bf D15} :996 (1977);\\
M.T. Grisaru and H.N. Pendleton, Nucl.\ Phys.\ {\bf B124}:81 (1977);\\
S.J. Parke and T. Taylor, Phys.\ Lett.\ {\bf B157}:81 (1985).


\end{thebibliography}
\end{document}